\documentclass[prl,aps,twocolumn]{revtex4}
\usepackage{epsfig} 
\usepackage{bm,amsmath}
\usepackage{subfigure}    

\usepackage[latin1]{inputenc}
\newcommand{\beq}{\begin{equation}}
\newcommand{\eeq}{\end{equation}}
\newcommand{\bea}{\begin{eqnarray}}
\newcommand{\eea}{\end{eqnarray}}
\begin{document}
  \title{Direct velocity measurement of a turbulent shear flow in a planar Couette cell.}
  \author{Michael J. Niebling$^{1}$, Ken Tore Tallakstad$^{1}$, Renaud Toussaint$^{2}$ and Knut J{\o}rgen M{\aa}l{\o}y$^{1}$}
 \affiliation{$^1$ Department of Physics, University of Oslo, P.O. Box 1048, 0316 Oslo, Norway}
  \affiliation{$^2$ Institut de Physique du Globe de Strasbourg, EOST / Universit\'{e} de Strasbourg, CNRS, 5 rue Descartes, F-67000 Strasbourg, France}
  \begin{abstract}
In a plane Couette cell a thin fluid layer consisting of water is sheared between a transparent band at Reynolds numbers ranging from 300 to 1400. The length of the cells flow channel is large compared to the film separation. To extract the flow velocity in the experiments a correlation image velocimetry (CIV) method is used on pictures recorded with a high speed camera. The flow is recorded at a resolution that allows to analyze flow patterns similar in size to the film separation. The fluid flow is then studied by calculating flow velocity autocorrelation functions. The turbulent pattern that arise on this scale above a critical Reynolds number of Re=360 display characteristic patterns that are proven with the calculated velocity autocorrelation functions. The patterns are metastable and reappear at different positions and times throughout the experiments. Typically these patterns are turbulent rolls which are elongated in the stream direction which is the direction the band is moving. Although the flow states are metastable they possess similarities to the steady Taylor vortices known to appear in circular Taylor Couette cells.
  \end{abstract}
  \maketitle
  \section{Introduction}
  \label{int}
Since the beginning of the 20th century the stability of fluid flows has been studied in unlike systems. Typical examples of such systems are natural convection in the atmosphere or convection in the earth mantle and the Rayleigh-B\'enard convection when a thin layer of fluid is heated from below \cite{Busseheat,Busseheat2}. When flow instabilities arise in these systems during the transition to turbulence often turbulent rolls appear that are also referred to as convection cells. \\
The shear flow of a thin fluid layer that is studied here also falls into this group of systems that display turbulent rolls often called Taylor vortices when the turbulent regime is reached.\\
The most prominent setup to study the instability of a sheared thin fluid layer is the circular Taylor Couette cell. In this setup two coaxial mounted cylinders with a small difference in radii rotate independently at different angular frequencies. The small difference in the cylinders radii leaves a thin gap that is filled with a fluid. Different angular frequencies cause a shear of the fluid layer. Centrifugal forces act on the fluid layer between the cylinder in this situation. This centrifugal forces depend on the radial position and on the angular frequencies of the cylinders. For some combinations of the angular frequencies and cylinder radii the resulting centrifugal forces give rise to an instability of the steady rotary flow. The instability leads to toroidal Taylor vortices or rolls that are again stable in time and space. These vortices are arranged in a periodic manner along the cylinders as described in \cite{landau} and studied in \cite{circ, circ2}.\\
In more recent experiments it was shown that Coriolis forces can as well lead to periodic flow patterns in shear flows as shown and discussed in \cite{tsuka}. In this work a planar Couette cell (see Fig.\,\ref{fig:expsetup} and \cite{bech, till}) was mounted on a rotating table with an angular frequency controlled independently from the shear velocity of the flow channel. The Coriolis force was then introduced by controlling the rotational frequency and direction of the table. These experiments show that the Coriolis force leads to a similar instability and patterns as the centrifugal force in the circular Taylor Couette cell. In fact in \cite{tsuka} a great variety of flow patterns emerged during the transition to the turbulent regime depending on the shear velocity and the Coriolis force. Weather the turbulent flow pattern are stable or meta stable in time and space at a fixed shear velocity depends on the direction and frequency of the table rotation. \\
In both above setups the most simple configuration of turbulent flow patterns are rolls that span through the whole flow channel in the stream direction. Such rolls are stacked on top of each other in a periodic manner where neighboring rolls rotate counter-wise and move in the opposite stream-wise direction. The streamlines in such a roll cell resemble a helix. The helices are stacked one above another in the stream-wise direction with alternating helix axis direction but same chirality. \\
The turbulent patterns originate from small perturbations of the flow that grow exponentially in time and lead to steady flow patterns under the above discussed conditions. When neither centrifugal nor Coriolis forces are present the turbulent patterns lose their stability and become meta stable. This is the case in the setup that is studied in the present work. \\ 
It is the main purpose of the present paper to quantify and describe the patterns that result as the stabilizing mechanisms of the fictitious forces are removed. We achieve this through a statistical analysis of the flow in terms of velocity correlation functions and velocity probability distributions. The resulting flow patterns are characterized by stream-wise vortices with alternating stream-wise flow direction between neighboring vortices.  \\ 
In simulations and analytical work of this setup by \cite{Nagata,Nagata2,bt, Tuckerman, Barkley, clever} meta stable elongated turbulent rolls have been reported and studied. In further investigations a low-order model is used to investigate self-sustaining rolls and effects of the boundary conditions in \cite{Waleffe,Waleffe2} and admissible fluids states and traveling wave solutions of the Navier-Stokes equation are simulated in \cite{cvitanovic, cvitanovic2, Cherhabili,Holstad,Holstad2}.\\
Coexistence of laminar and turbulent regions where investigated numerically by \cite{Duguet,Duguet2,Rolland}. Further the flow in a plane Couette cell was characterized into four flow regimes by \cite{Tuckerman, Barkley} using direct numerical simulations. In preceding experiments of the planar Taylor Couette cell a rheoscopic (aka kalliroscopic) fluid was sheared and pictures of the patterns taken \cite{JJH,KJM,TurbulentSpots}. Fluid injection into the flow creates turbulent spots which were studied in detail by \cite{TurbulentSpots, Daviaud, Duguet, Duguet2}. In addition in \cite {Bottin4, Dauchot, Bottin} the flow patterns in a planar Couette cell close to a wire spanning the cell parallel to the walls were investigated experimentally by the use of different techniques such as laser Doppler velocimetry and rheoscopic particles. When the Reynolds number is used as the control parameter and suddenly decreased in quench experiments the decay time of vortex structures have been measured in experiments by \cite{Bottin, Bottin2} and experiments are compared to numerical simulations in \cite{Bottin3}. The interaction between thermal convection and shear flows in a plane Couette cell is studied through numerical simulations in \cite{Busse, Busse2}.\\ Using these method periodic and elongated flow structures along the stream direction could be visualized and studied. Compared to Taylor vortices found in circular cells these patterns in a planar Couette cell are much shorter in the stream direction and fluctuate in time and space.\\  
To achieve a direct measurement of the flow velocity we decided to use a correlation image velocimetry technique (CIV). For this purpose tracer particles with the same mass density as the sheared fluid are added to the fluid. Adding density matched tracer particles is a standard non invasive technique. This technique allows a direct measurement of the fluid velocity that reveals additional characteristic features of the flow patterns which were not accessible using the previous techniques where only the flow direction was visualized \cite{tsuka,JJH,KJM,TurbulentSpots}. \\
In more recent studies 3d particle tracking velocimetry (PTV) \cite{Hagiwara,Krug,Luthi,Hoyer} was applied to flow experiments in a plane Couette cell and the flow field measured in three dimensions for the first time.  
In the following the relative arrangement and shape of the appearing flow patterns is further discussed and quantified through velocity autocorrelation functions.
\section{Experimental Setup}
\label{expsetup}
\subsection{Mechanical setup}
\begin{figure}[ht!]
  \includegraphics[width=1.0\columnwidth]{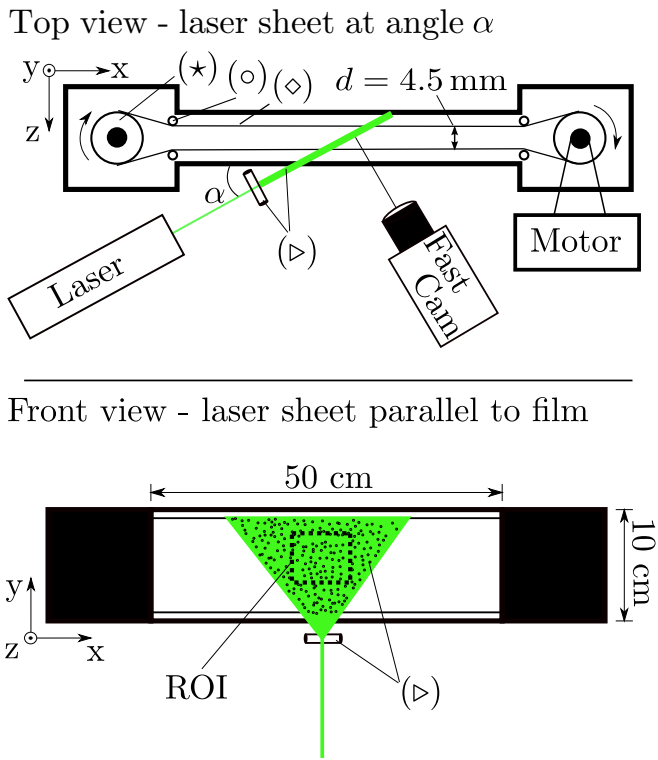} 
  \caption{\label{fig:expsetup} (\textbf{Upper panel}) Top view of the planar Couette Cell. A mylar film ($\diamond$) is put around two cylinders ($\star$), and set in rotation by a motor. Additional four smaller cylinders ($\circ$) set the internal gap spacing between the inner planes of the film, $d$. The entire setup is placed inside a waterproof container with geometry as indicated by the thick solid boundary. The thinner channel in the middle is made of Plexiglas. A laser sheet ($\triangleright$) is used to illuminate a chosen cross section of the flow channel. In this particular illustration, the laser sheet enters the channel at an angle $\alpha$ as indicated (cf. Fig.\,\ref{fig:expconfigs}). (\textbf{Lower panel}) Front view of the Couette Cell. In this particular illustration, the laser sheet ($\triangleright$) enters the cell from below, and spans the $xy$-plane. The fluid is seeded with density matched fluorescent particles. Indicated by a dashed line is the $4\, \mathrm{cm}\times4$\,cm Region Of Interest (ROI), captured by a fast camera in order to follow the fluid motion (cf. Fig.\,\ref{fig:expconfigs}).}
\end{figure}

The planar Couette Cell is shown in Fig.\,\ref{fig:expsetup}. It consists of a confining volume ($\sim2.5\,$liters), in which a thin (thickness $\sim50\,\mu$m) Mylar belt ($\diamond$) is rotating around two cylinders ($\star$). The left cylinder can be moved in the $x$-direction so as to tighten the belt, whereas the right cylinder is driven by a DC motor. Velocities in the interval $(0-60)\,$cm/s of the Mylar film can be obtained. Four small steering cylinders ($\circ$) are mounted, two at each end of a narrow flow channel. These cylinders are fixed and define the inner spacing between the Mylar film. In the narrow flow channel, the film has a height of $10\,$cm in the $y$-direction, an inner separation distance $d=4.5\,$mm in the $z$-direction, over a length of $50\,$cm along the $x$-direction. Furthermore, the flow channel is made of Plexiglas to ensure light transparency.

The Mylar film was sealed as a belt, with a joint without overlap. 
The boundary conditions are open to the rolls on the sides and closed to top and bottom of the cell. Turbulence arises first at the sides of the cell due to the noise induced by the rolls as shown with rheoscopic particles during experiments in the same cell by \cite{KJM}.  \\

The initial preparation of the Couette Cell is done by the following routine: 1) the cell is filled with water, 2) three drops of liquid soap are added only to prevent particle aggregation and clustering by reducing the surface tension (this has no effect on the experiments otherwise), 3) $7$ grams of fluorescent polyethylene tracer particles (Cospheric Fluorescent Red) with diameter $125-150\,\mu$m are added to the water. These particles are density matched with water ($0.995\,$g/ml). Furthermore, all experiments are performed in room temperature at roughly $~22^\circ$C, and constant humidity. 
 
When the film is set in motion, the fluid/particle mixture between the two sides of the film, moving in opposite $x$-directions, starts to flow. We define the absolute value of the film speed ($x$-direction) as $U$. Depending on the value of $U$, the flow will be either laminar or turbulent. We emphasize that this flow geometry, gives an overall zero average flow velocity. 

\subsection{Optical setup}
The tracer particles added to the water are fluorescent 125-150 $\mu$m polyurethane microshperes. Thus in order to visualize the flow, we use a beam from a Lexel 95-3 argon ion laser of wavelength $514.5\,$nm, at up to $3.0\,$W. The beam is sent through a cylindrical lens, creating a laser sheet (see Fig.\,\ref{fig:expsetup}) of thickness $L_d\sim1.0\,$mm. As shown in the upper and lower panel of Fig.\,\ref{fig:expsetup}, the laser sheet ($\triangleright$) can be positioned differently, in order to span different cross sections of the flow. When the laser light meets the tracer particles, they glow in a red color and allow to visualize the flow. The intensity profile of the laser is Gaussian, and thus, the particles scattered intensity will vary accordingly. The particles are only visible as long as they are positioned inside this light intensity profile, resulting in a two dimensional visualization of the three dimensional flow. 
To record the flow, a fast camera (Photron Fastcam SA5) is positioned normal to the laser sheet, as shown in the upper panel of Fig.\,\ref{fig:expsetup}. The recording rate of the camera is $2000\,$ frames per second (fps), at a spatial resolution of $40\,\mu$m/pixel, and roughly $10\,000$ images are acquired per experiment. 
The velocity of the Mylar film $U$ is the only tuning parameter, and we define the Reynolds number as
\begin{align}
 \text{Re}=\frac{U}{\nu}\frac{d}{2}\ ,
\end{align}
where $\nu=1.0$\, $\mathrm{mm}^2/\mathrm{s}$ is the kinematic viscosity of water, and $d/2$ is half the separation distance between the two sides of the Mylar film. Previous studies have shown \cite{till,KJM} that the critical Reynolds number, i.e where the flow goes from laminar to turbulent, is $\text{Re}_c\approx 360 \pm 20$.

\begin{figure}[ht!]
  \includegraphics[width=1.0\columnwidth]{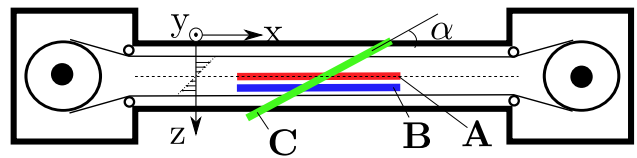} 
  \caption{\label{fig:expconfigs} Our experiments are grouped into three sets, depending on the orientation of the laser sheet. In \textbf{set A} the laser sheet spans the \textit{xy}-plane, in the center of the model, at a distance $2.3\,$mm from the moving front film. In \textbf{set B} the laser sheet spans the \textit{xy}-plane, but now further towards the front of the flow channel, at a distance $1.3\,$mm from the moving front film. In \textbf{set C} the laser sheet enters the cell at an angle $\alpha=52^\circ$, spanning an $x'y'$-plane as defined in Eq.\,(\ref{eq:coordtrans}). Also indicated along the $z$-axis is the linear velocity gradient in the case of a low-Re, laminar flow.}
\end{figure}

In this study, we have performed three sets of seven experiments ($21$ in total), at roughly the same range of $\text{Re}\approx\{300,400,460,610,860,1120,1400\}\pm$20. The $\text{Re}=300$ experiments are laminar, whereas the rest is in the turbulent regime. For each experimental set, the orientation of the laser sheet has been changed, so as to probe different cross sections of the flow. The coordinate system is shown in the upper and lower panel of Fig.\,\ref{fig:expsetup}. We define $x$ as the \textit{stream-wise} direction, $y$ as the \textit{span-wise} direction, and $z$ as the \textit{wall-normal} direction. A rotation of an angle $\alpha$, of the original coordinate system, around the $y$-axis, gives the transformed coordinates ($x',y',z'$):
\begin{align}
x'&=x\cos{\alpha}-z\sin{\alpha}\notag\\
y'&=y\label{eq:coordtrans}\\ 
z'&=x\sin{\alpha}+z\cos{\alpha}\notag\ .
\end{align}
The different orientations of the laser sheet are shown in Fig.\,\ref{fig:expconfigs}. In set A, the laser sheet spans the $xy$-plane in the center of the cell, at a distance $d/2\approx 2.3\,$mm from the moving front film. In set B, the laser sheet spans the $xy$-plane further towards the front of the flow channel, at a distance $1.3\,$mm from the moving front film. Set A and B have identical camera position and experimental alignment apart from the position of the laser sheet.   
To distinguish between these two sets of experiments, from now on, variables will be denoted by a subscript $_c$ and $_f$ for center view (set A) and front view (set B) respectively. In both sets A and B, the images acquired spans a $4\, \mathrm{cm}\times 4\,$cm square in the laser sheet, as shown in the lower panel of Fig.\,\ref{fig:expsetup}, and is the region of interest (ROI) (see also videos \cite{supp}). In set C, the laser sheet is rotated around the $y$-axis, at an angle $\alpha=52^\circ$ from the $x$-axis, and spans now an $x'y'$-plane as defined in Eq.\,(\ref{eq:coordtrans}). Coordinates in this plane will from now on be referred to with the superscript $'$. The ROI in set C, spans a $0.55\,\mathrm{cm}\times4\,$cm rectangle of the flow cell (see also videos \cite{supp}).  
\subsection{Analysis procedure}
\label{anproc}
To obtain the velocity field of the flow, each sequence of images is analyzed using a standard multi-pass \textit{correlation image velocimetry} (CIV) technique \cite{piv,nieb1}. A presision test of this method was done in i.e. \cite{nieb1}. The pictures are recorded at a resolution of 1024x1024 pixels. Using the CIV technique the velocity fields are calculated at a lower resolution than the initial gray scale pictures which are recorded at a resolution of 256 pixels/cm. The final resolution of the velocity field is 16 measurement points per cm in set A and B and 30 measurement points per cm in set C. This resolution is carefully chosen to fit for a statistical analysis of vortices similar in size to the plate separation $d$. \\
The finite thickness of the laser sheet illuminates particles within a distance interval normal to the region of interest (ROI) of approximately 1mm. The observed ROI is therefore initially 3 dimensional with the dimension 4cm by 4cm by 1mm. Particles can be tracked easier in a thicker laser sheet since they stay illuminated longer even though they move somewhat normal to the ROI in the wall normal $z$-direction. 
When the pictures are taken the normal dimension of the recorded volume is projected on a 2 dimensional plane. With a thick laser sheet particles that appear at the same position in the pictures can have different $z$-coordinates and therefore different velocities in the recorded 2d pictures. For the correlation tracking the local particle constellation must be related between two subsequent pictures. It is therefore important that the laser sheet is not too thin that particles get lost too fast due to movement normal to the film. Neither should the laser sheet be too thick causing the recorded particles to locally have very different velocity and destroying the local constellation of particles.\\  
\begin{figure*}[th!]
  \centerline{
  \subfigure[]{
    \epsfig{height=.29\textwidth,file=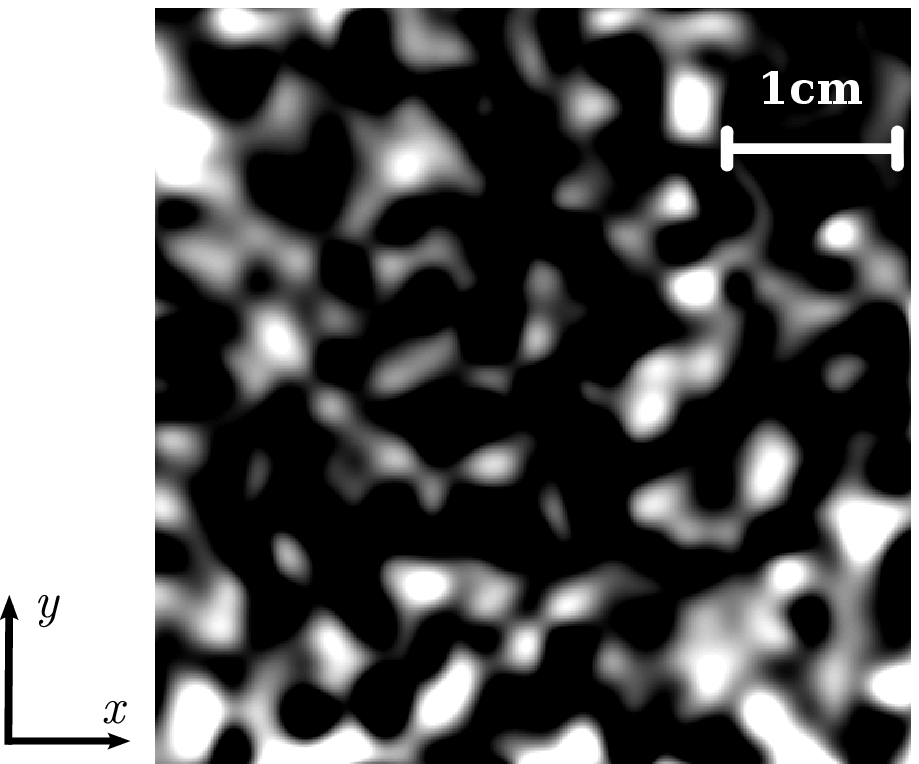}
    \label{r1}
  }
 \subfigure[]{
    \epsfig{height=.29\textwidth,file=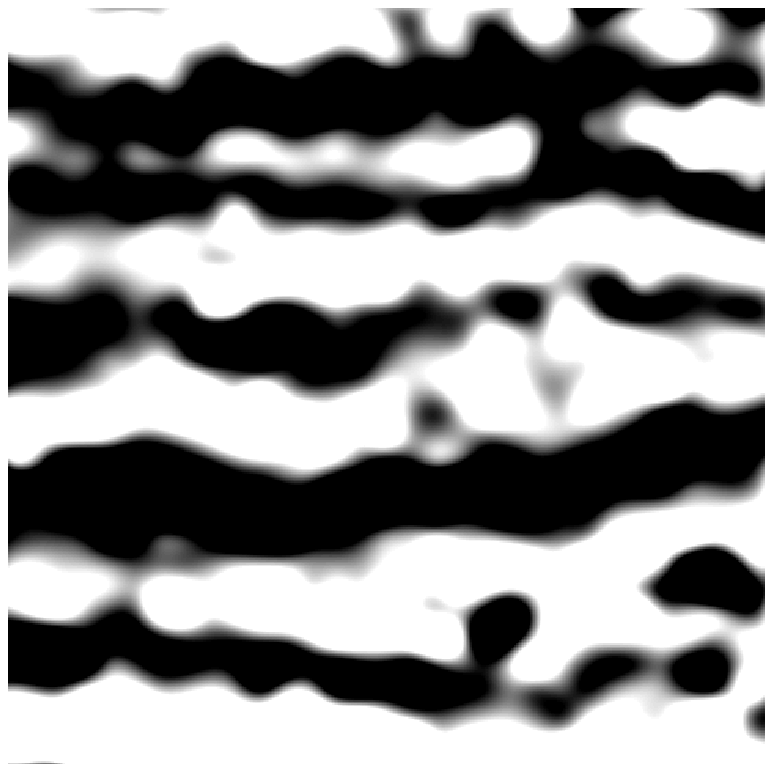}
    \label{r2}
  }
  \subfigure[]{
    \epsfig{height=.29\textwidth,file=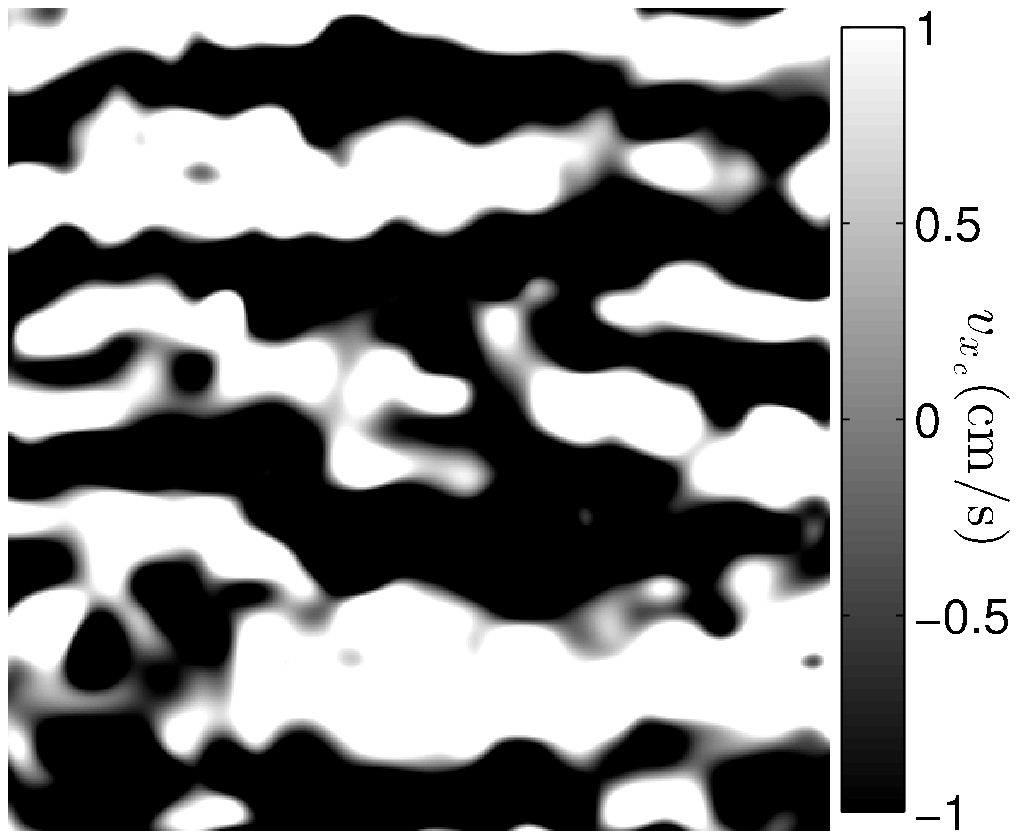}
    \label{r3}
   }}  
  \caption[Optional caption for list of figures]{Snapshots of the horizontal velocity $v_{x_c}$-component in experiment set A display typical flow fields in the $x_cy_c$-plane of 4\,cm x 4\,cm ROI area at $2.3\,$mm from the moving front film.. Reynolds numbers are Re=300 (laminar) in \subref{r1}, 400 in \subref{r2} and 1120 in \subref{r3}. To enhance the contrast velocities are represented by a linear gray scale only from -1cm/s (black) up to 1cm/s (white) although the actual range of velocity values is much larger. Values below or above this linear scale are represented black or white respectively. For smoother pictures the spatial resolution was increased by linear interpolation. Stripes of alternating flow direction are visible that are layered on top of each other.}
  \label{rall}
\end{figure*}
\begin{figure*}
  \centerline{
  \subfigure[]{
    \epsfig{height=.29\textwidth,file=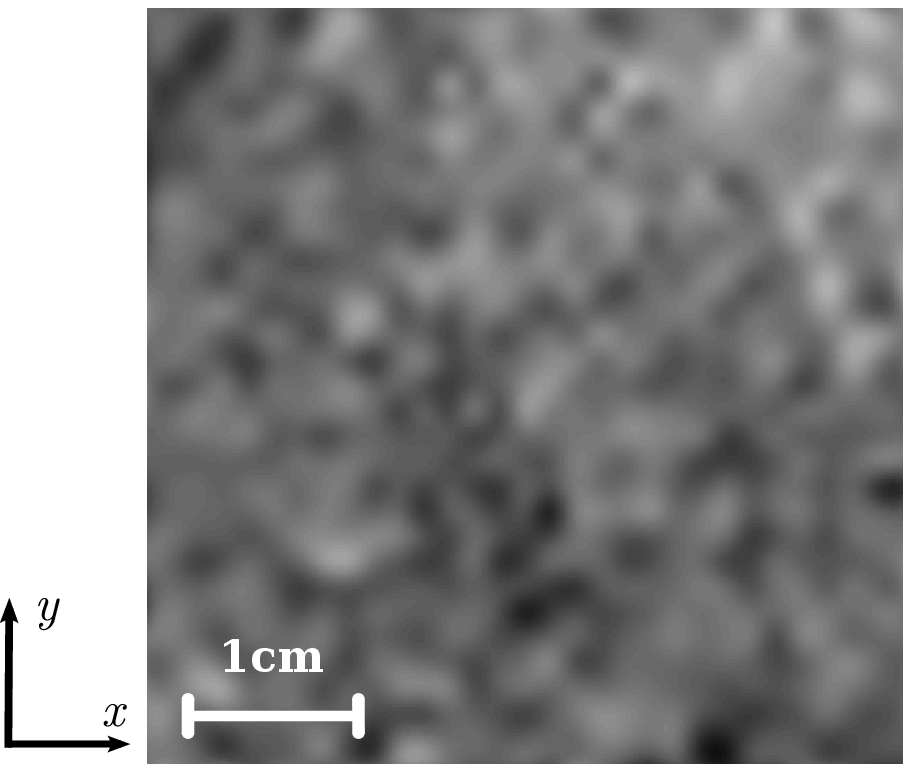}
    \label{w1}
  }
 \subfigure[]{
    \epsfig{height=.29\textwidth,file=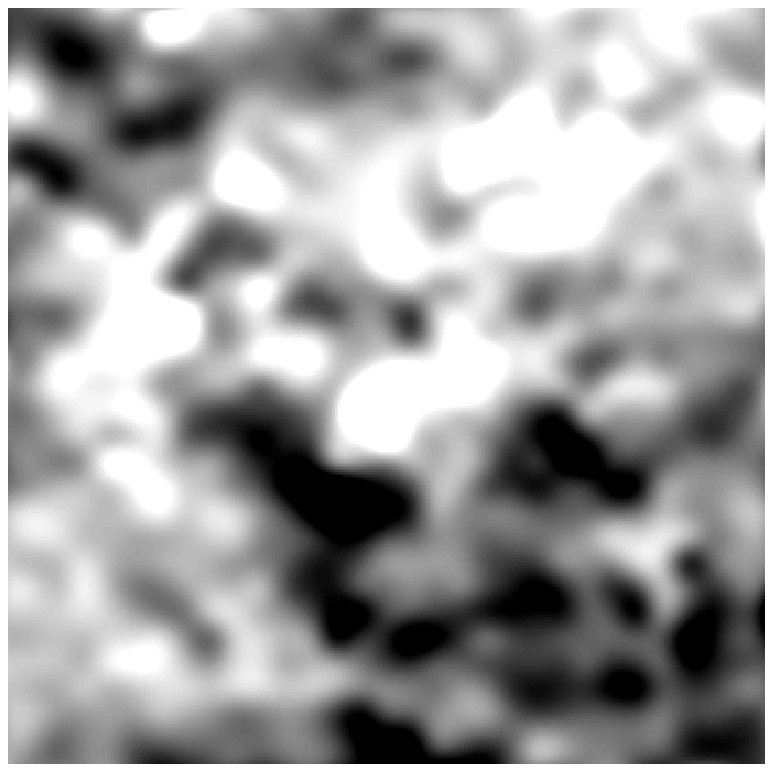}
    \label{w2}
  }
  \subfigure[]{
    \epsfig{height=.29\textwidth,file=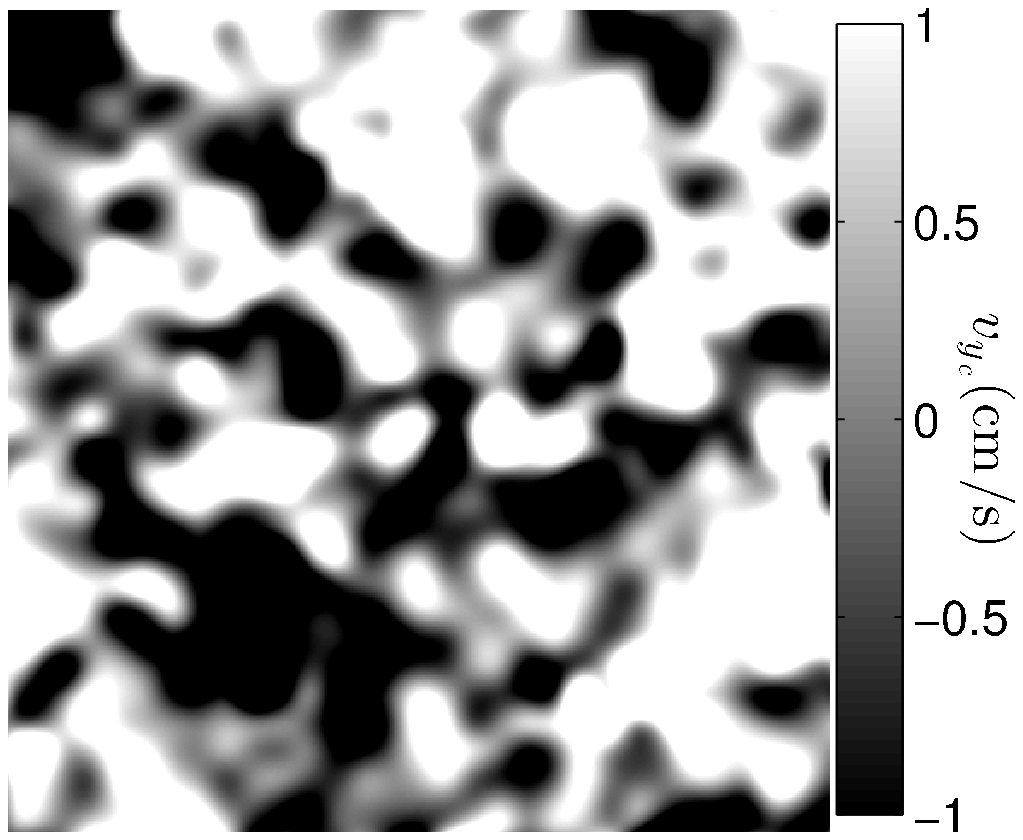}
    \label{w3}
   }}  
  \caption[Optional caption for list of figures]{Corresponding to Fig. \ref{rall} in all other specifications. Snapshots of the vertical velocity $v_{y_c}$-component in experiment set A. The snapshots show no obvious flow patterns.} 
  \label{wall}
\end{figure*}
\section{Results}
\label{Results}
\subsection{Front View}
\begin{figure*}
  \centerline{
  \subfigure[]{
    \epsfig{height=.34\textwidth,file=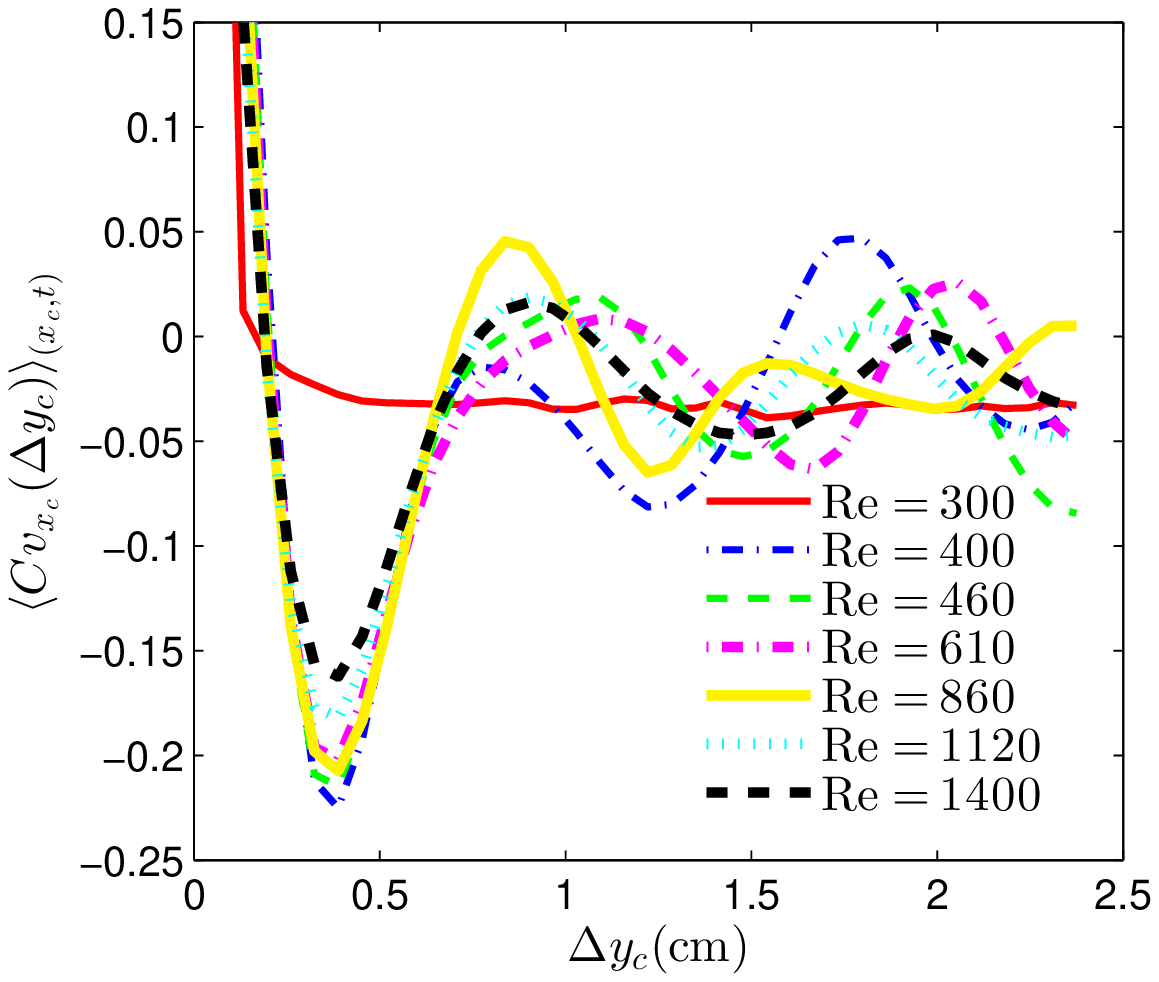}
    \label{FVA1}
  }
 \subfigure[]{
   \epsfig{height=.34\textwidth,file=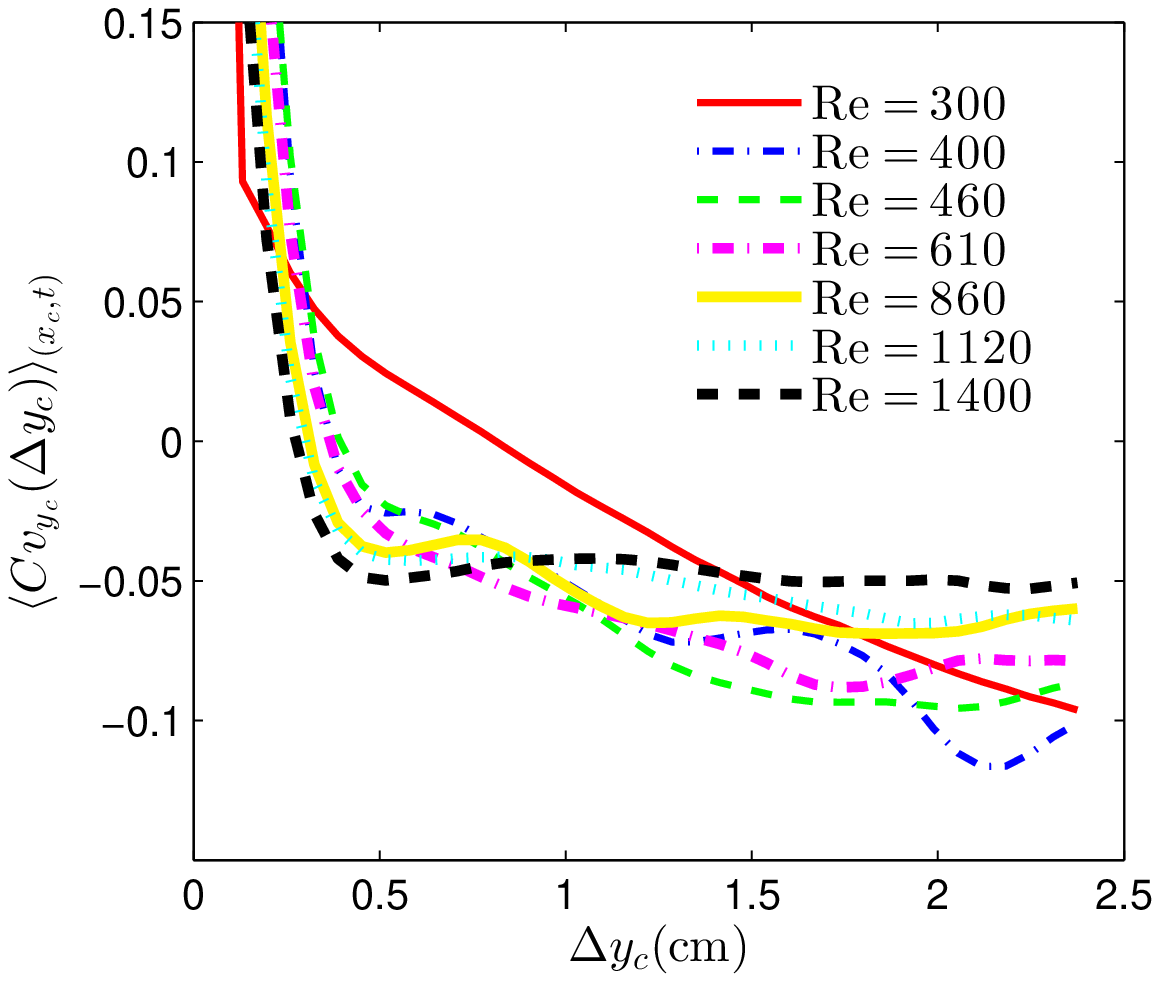}
    \label{FVA2}
 }}
  \centerline{
  \subfigure[]{
    \epsfig{height=.34\textwidth,file=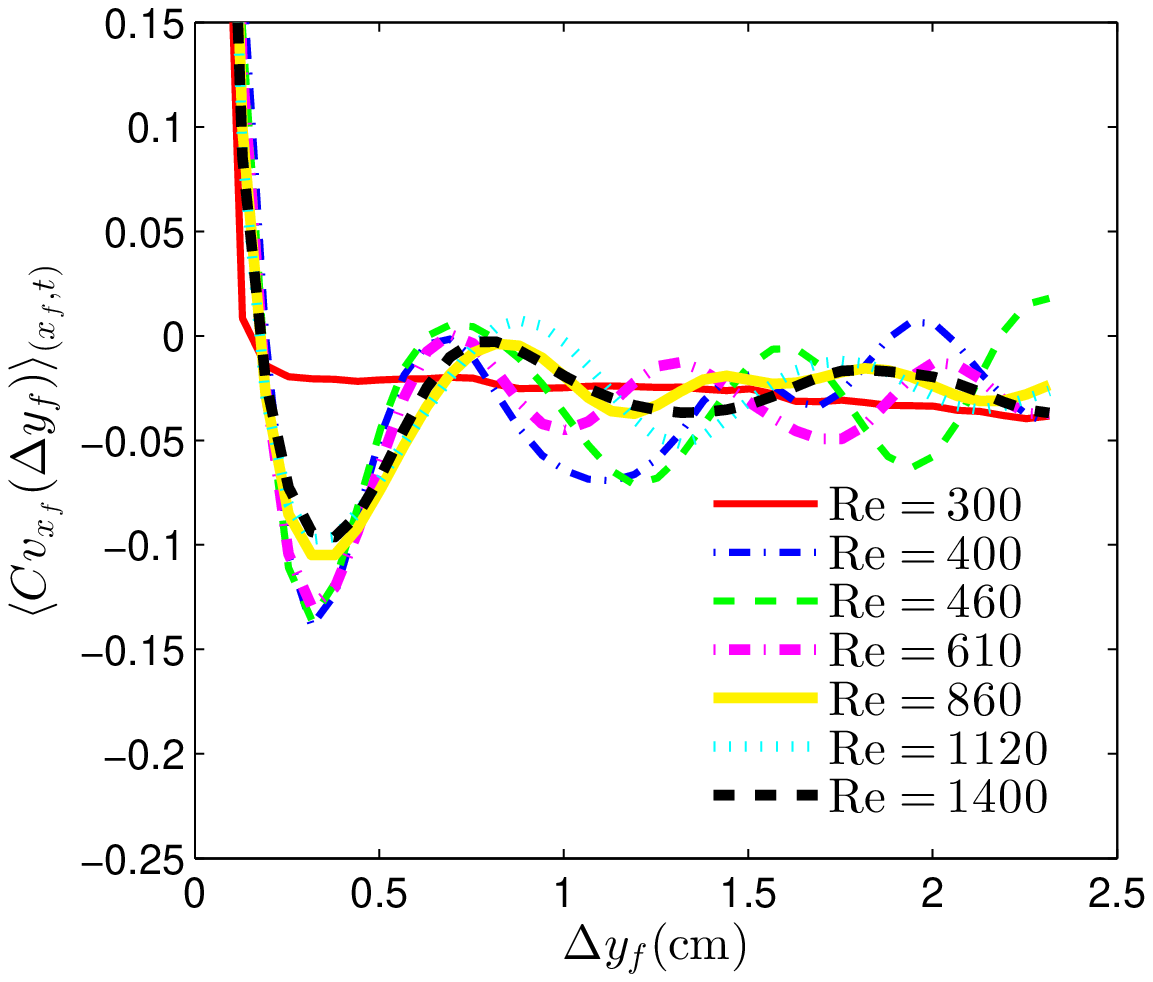}
    \label{FVA3}
  }
 \subfigure[]{
    \epsfig{height=.34\textwidth,file=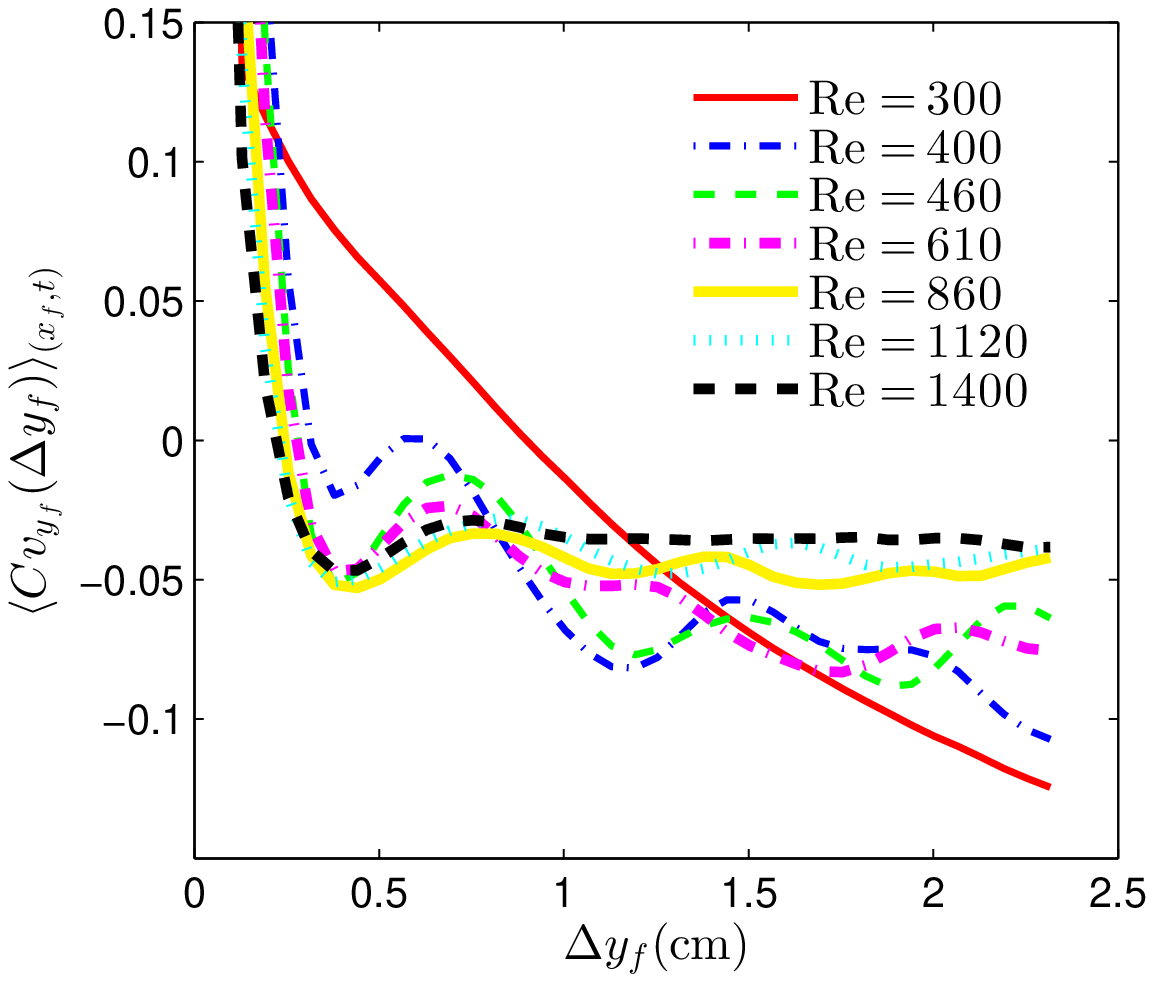}
    \label{FVA4}
  }} 
 \centerline{
  \subfigure[]{
    \epsfig{height=.34\textwidth,file=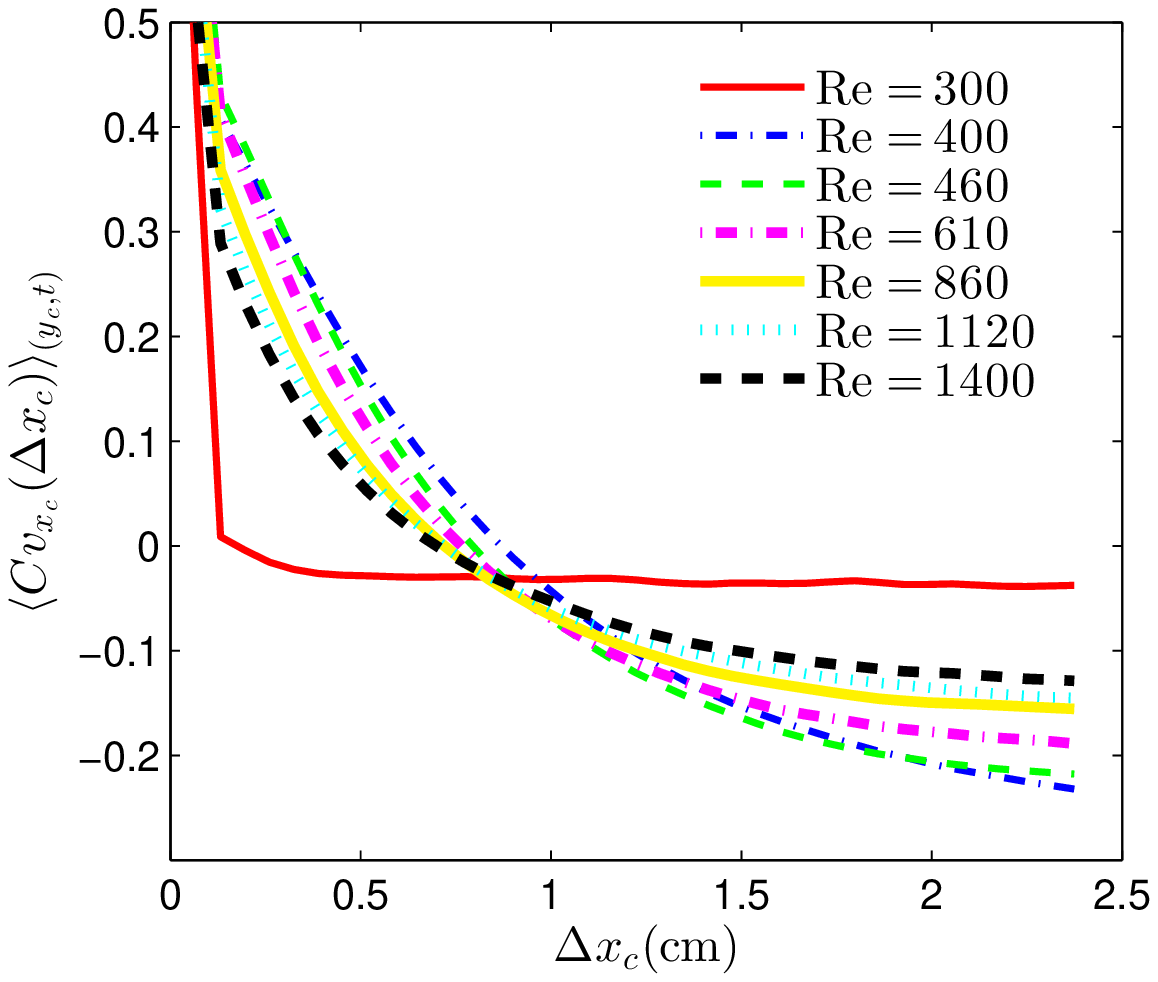}
    \label{FVA5a}
  }
 \subfigure[]{
   \epsfig{height=.34\textwidth,file=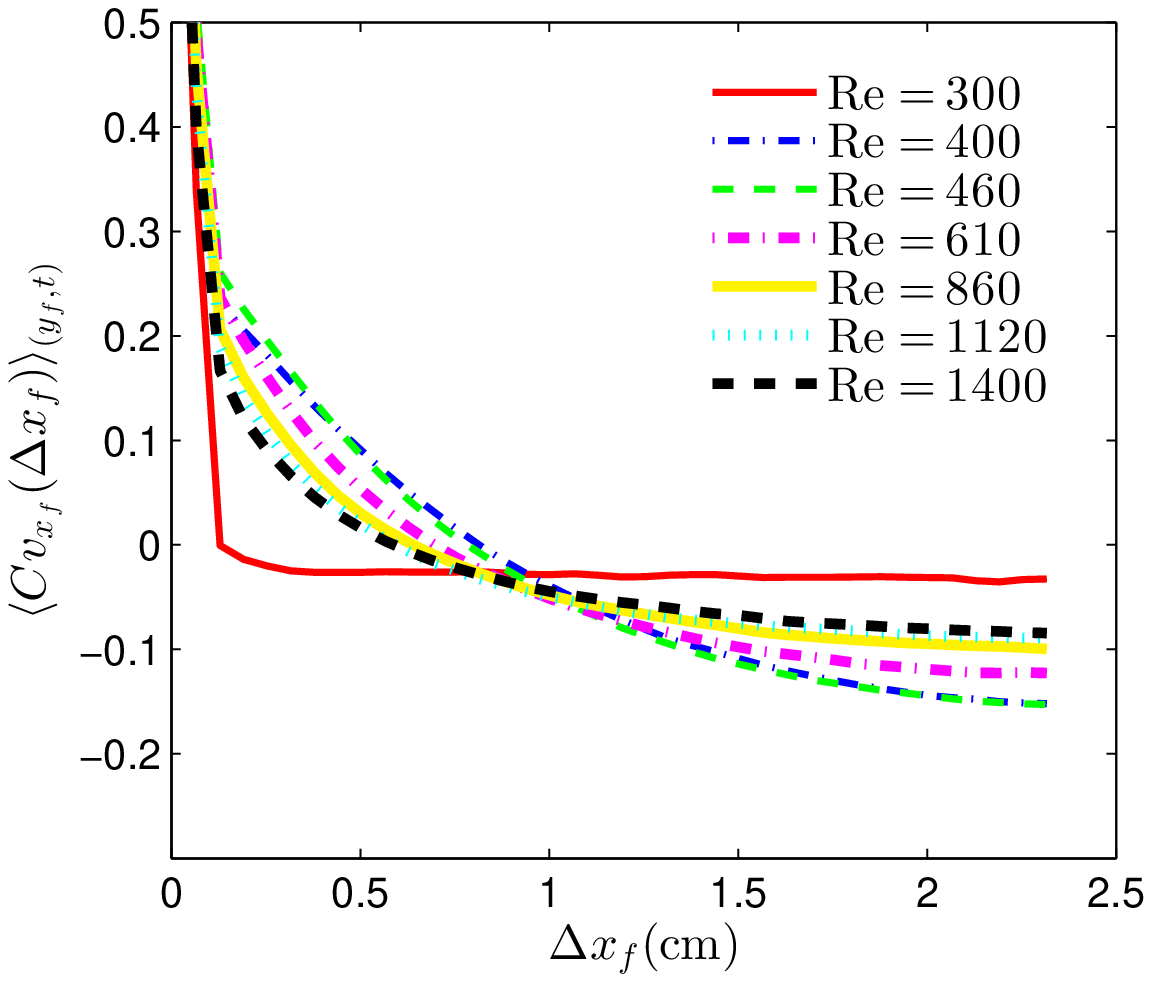}
    \label{FVA6a}
  }} 
  \caption[Optional caption for list of figures]{Velocity autocorrelation as a function of the separation in $y$-direction: $\Delta y$. \subref{FVA1} shows the autocorrelation of the $v_{x_c}$-component and \subref{FVA2} of the $v_{y_c}$-component in the middle of the film at a position 2.3mm from the front film. \subref{FVA3} shows the autocorrelation of the $v_{x_f}$-component and \subref{FVA4} of the $v_{y_f}$-component close at a position 1.3\,mm from the front film. \subref{FVA5a} and \subref{FVA6a} show velocity autocorrelation functions of the $v_{x}$-component as a function of the separation in $x$-direction: $\Delta x$. \subref{FVA5} shows the autocorrelation functions gathered from experiments with the laser sheet in the middle of the film at a position 2.3\,mm from the front film. \subref{FVA6} shows autocorrelation functions from experiments with the laser sheet at position 1.3\,mm from the front film.}
  \label{FVAall4}
\end{figure*}
\begin{figure*}
  \centerline{
  \subfigure[]{
    \epsfig{height=.34\textwidth,file=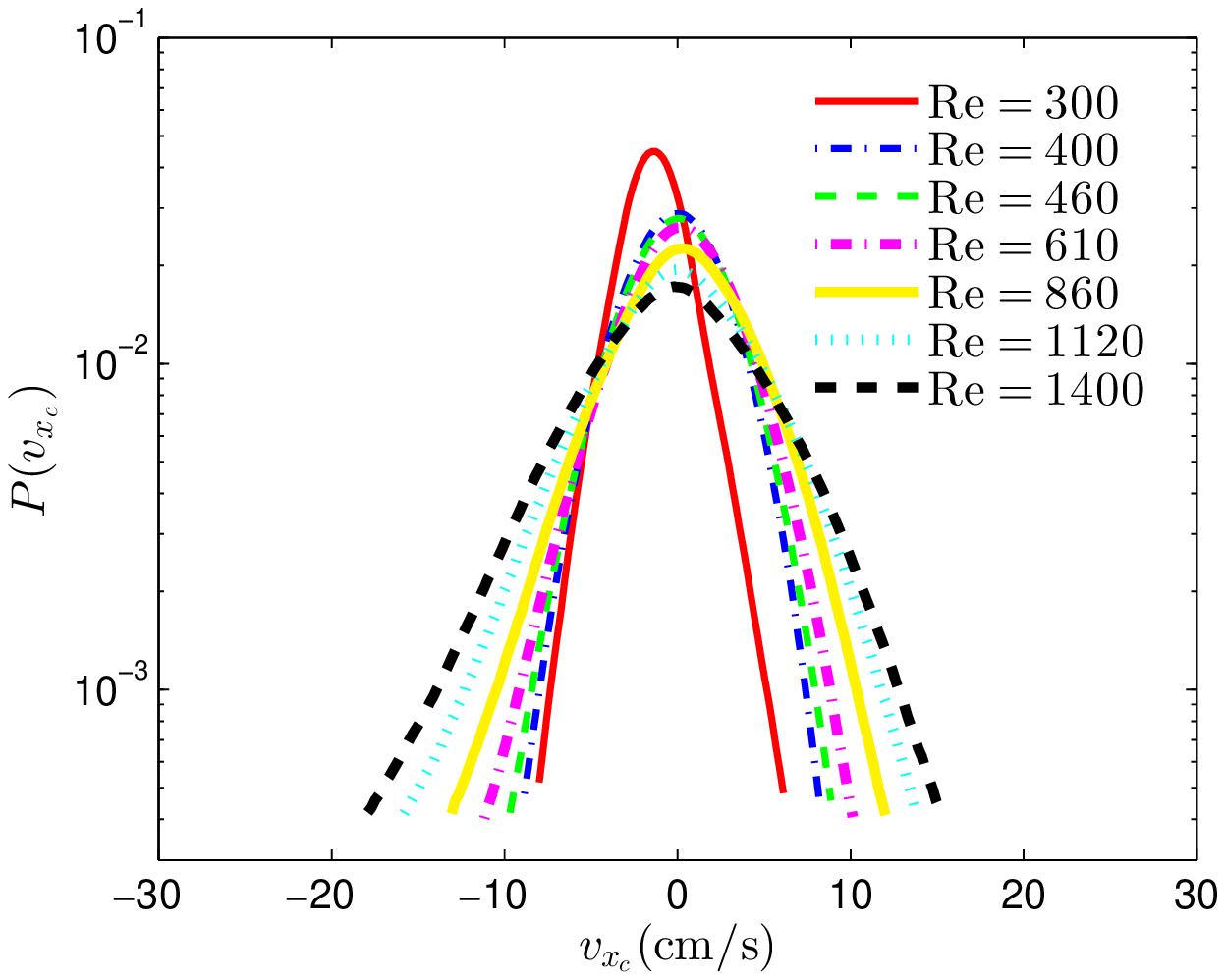}
    \label{FVA5}
  }
 \subfigure[]{
    \epsfig{height=.34\textwidth,file=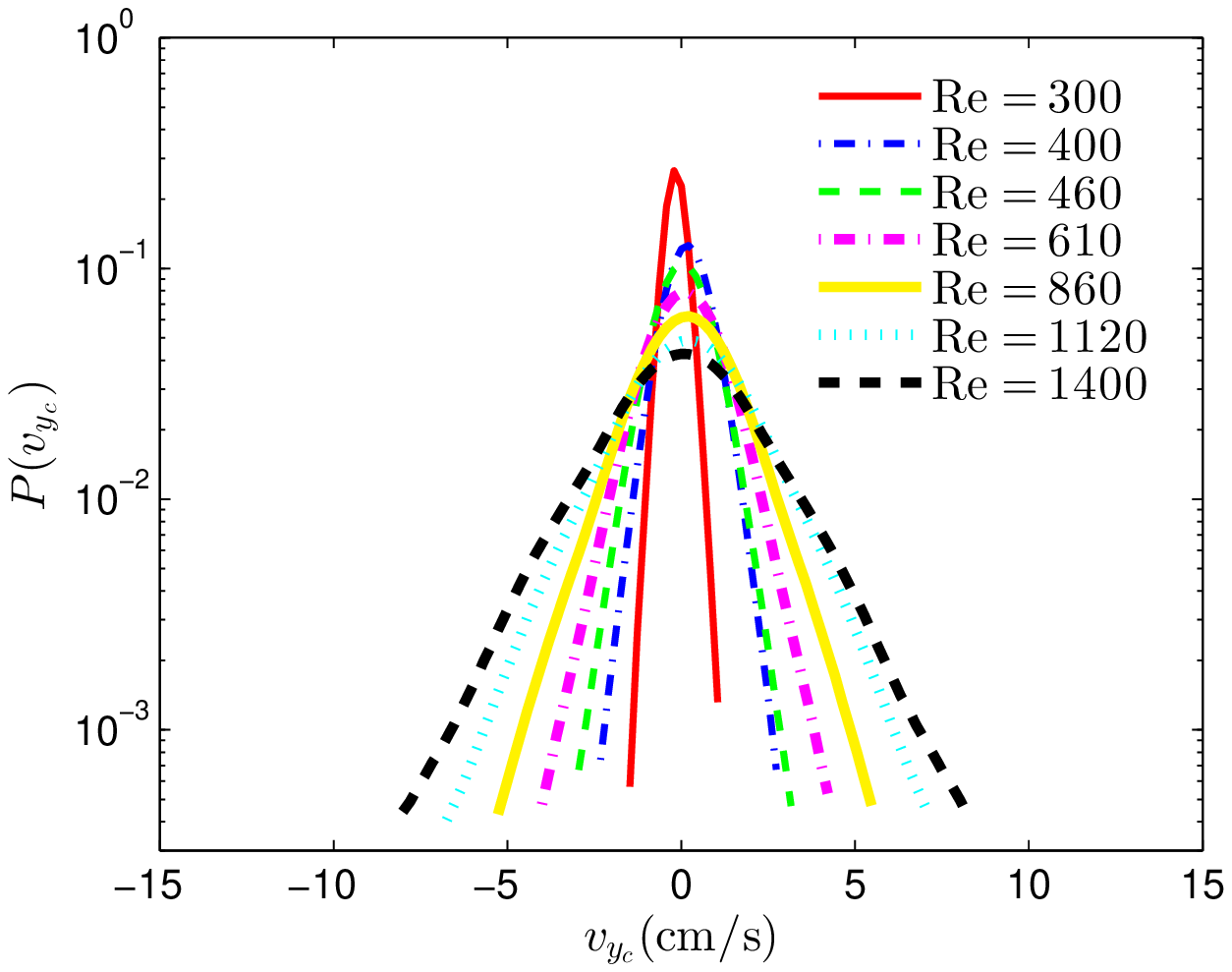}
    \label{FVA6}
  }}  
   \centerline{
  \subfigure[]{
    \epsfig{height=.34\textwidth,file=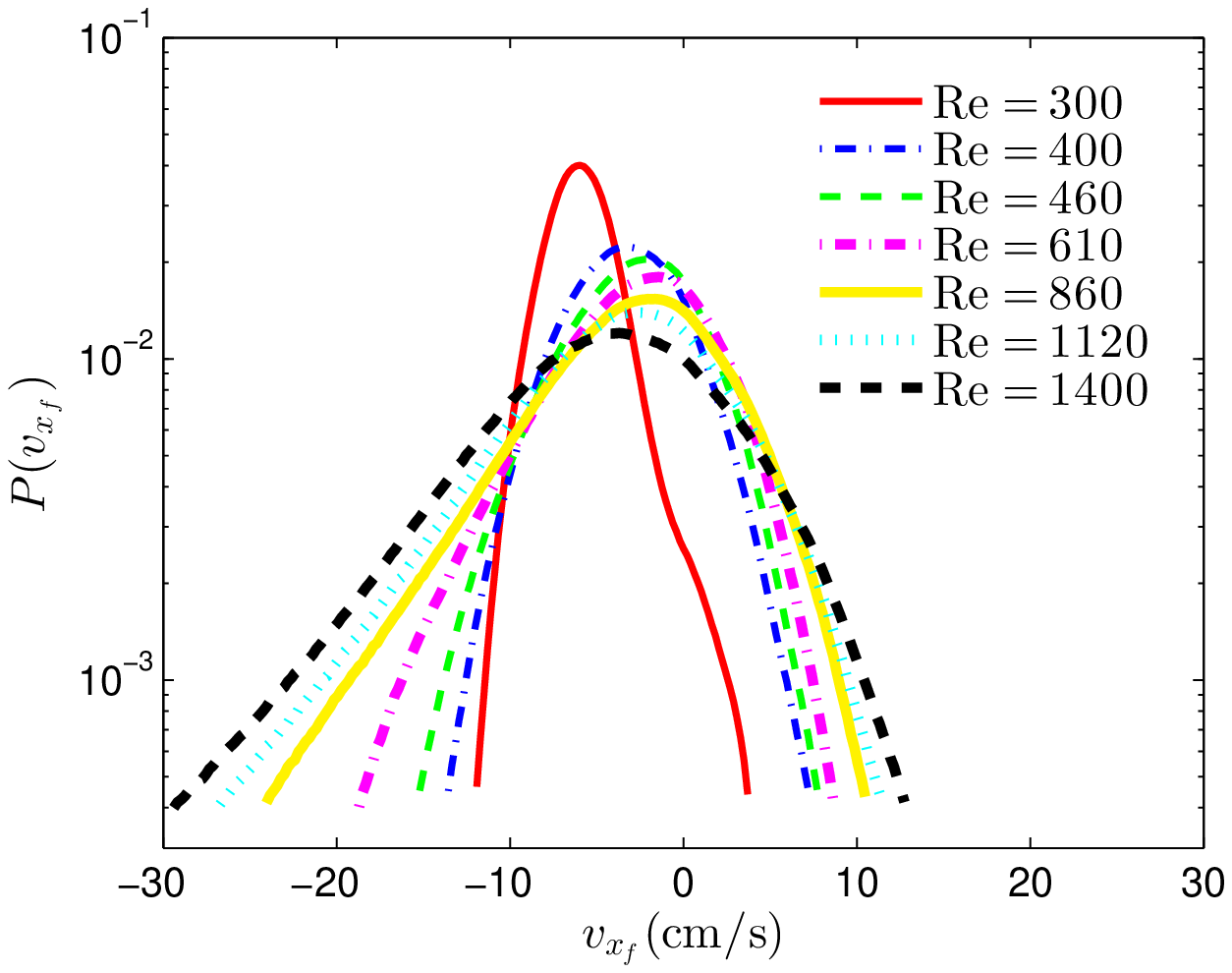}
    \label{FVA8}
  }
 \subfigure[]{
    \epsfig{height=.34\textwidth,file=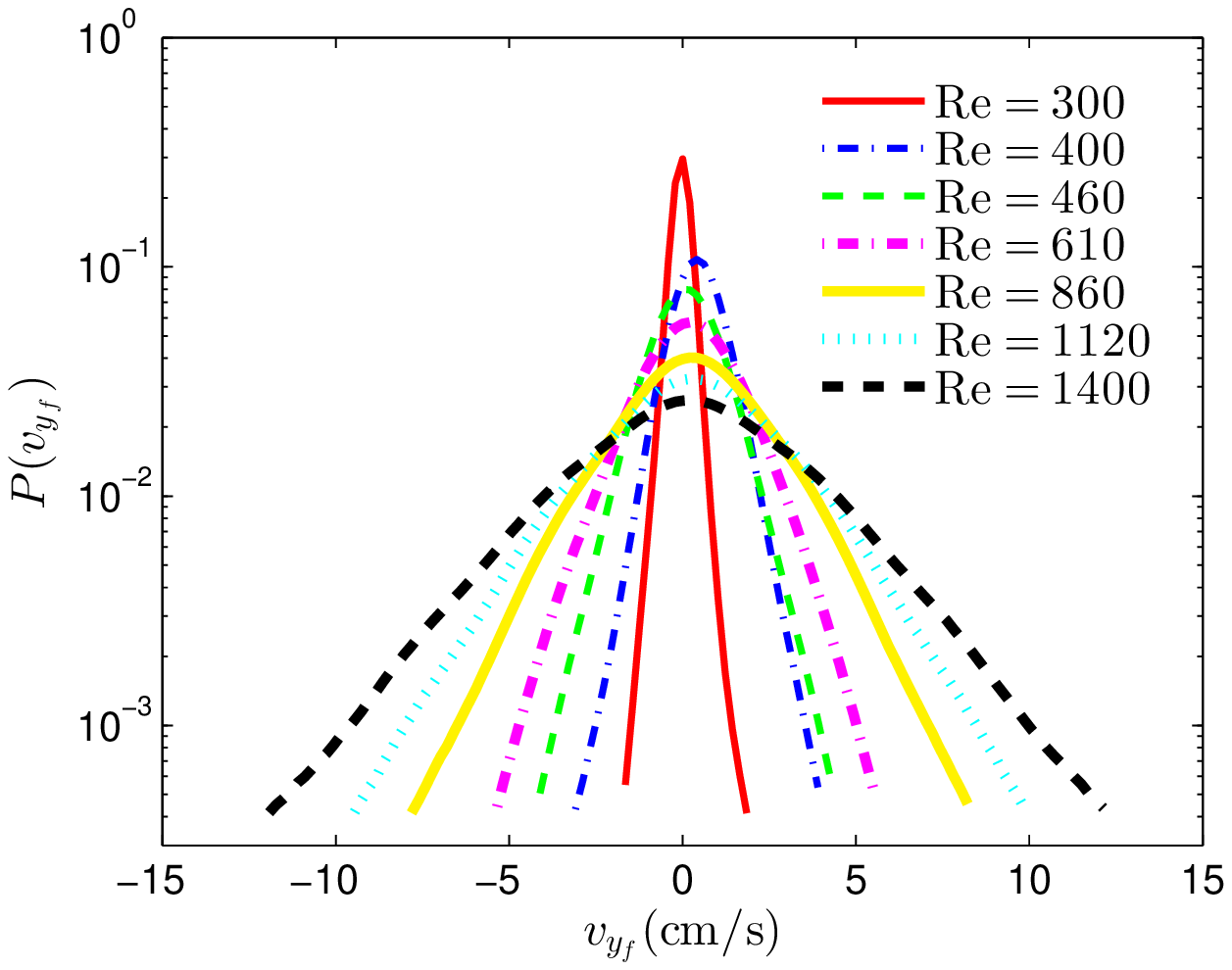}
    \label{FVA9}
  }}  
  \caption[Optional caption for list of figures]{Velocity distributions of the two experimental sets A \subref{FVA5}-\subref{FVA6} in the middle between the film and B \subref{FVA8}-\subref{FVA9} close to the front film. \subref{FVA5} and \subref{FVA8} show distributions of the $v_x$ component. \subref{FVA6} and \subref{FVA9} show distributions of the $v_y$ component. All distributions are time and space averaged.}
  \label{FVAall3}
\end{figure*}
In example Figs.\,\ref{rall} and \ref{wall} three grays scale snapshots of the resulting flow profiles are shown for 3 different Reynolds numbers. All snapshots are recorded in the middle between the two sides of the film and from experiment set A. The gray scale corresponds to a velocity range from -1cm/s (black) to 1cm/s (white). Velocity values are visualized by a linear gray scale inside this range and values below or above this linear scale are represented black or white respectively. Actual values of the flow velocity are much higher in the turbulent cases but this range qualitatively highlights the vortex structure in the flow and increases the contrast. The spatial resolution in these snapshots is higher than the velocity data that was calculated by the CIV technique: this resolution is achieved by linear interpolation to improve the readability of the snapshots. \\
Figs. \ref{r1}-\ref{r3} show the $v_{x_c}$ component and the $v_{y_c}$ component is shown in Figs. \ref{w1}-\ref{w3}. Figs. \ref{r1} and \ref{w1} show snapshots of the laminar flow at Reynolds number of Re=300. The next two snapshots in Fig. \ref{r2}-\ref{r3} and \ref{w2}-\ref{w3} are taken at a higher Reynolds number that leads to a turbulent flow Re=400 and Re=1140. \\
For Re=300 the particles move randomly left or right depending on slight differences in their wall-normal $z$-position. In direct observations particle trajectories in the laminar regime Re=300 are straight horizontal lines that span through the whole ROI. In contrast to the laminar case the turbulent flow displays horizontal layers or clusters of alternating coherent $v_{x_c}$-direction (see Figs. \ref{r2}-\ref{r3} ). The alternating layers are elongated in the $x$-direction and the flow velocity is much higher as can be also seen in the velocity probability distribution that are shown further down. Direct observations further shows particle trajectories are now strongly curved.\\
The extension of these layers in the $y$-direction is around 4\,mm before the $v_{x_c}$ direction is reversed. This separation in $y$-direction is much shorter than the stream-wise length of the coherent clusters of $v_x$ in $x$-direction. The separation is however comparable to the film separation $d$. In stream-wise direction these clusters or layers can be much larger than the observed ROI of 4\,cm. Comparing Fig. \ref{r2} to \ref{r3} we got the impression that the length of the coherent $v_x$-clusters decreases in $x$-direction while the Reynolds number is increased. This is an observation that later will be confirmed by velocity correlation functions. The distance in $y$-direction until the flow $v_x$-direction is reversed however does not change noticeably in the snapshot with the Reynolds number Fig. \ref{r3}. These observations will also be shown by velocity correlation functions in the next paragraphs.\\
The snapshots of the $v_{y_c}$-velocity component show much less order. The $v_{y_c}$-component in the laminar snapshot Fig. \ref{w1} has small values between -1cm/s and 1cm/s with no obvious order. At a turbulent Reynolds number the $v_y$ component in Figs. \ref{w2} and \ref{w3} displays structures that appear to have a typical size but seem to be isotropic and randomly distributed. The stream patterns in the $v_{x_c}$ and $v_{y_c}$-component are constantly changing and reappear at different times at different positions.\\
In the following analysis autocorrelation functions and velocity distributions are calculated using data directly obtained by the CIV method to further quantify the flow patterns as seen in Figs. \ref{r2} and \ref{r3}. The velocity autocorrelation functions are calculated by
\begin{equation}
\label{autoc}
Cv_{\alpha}(\Delta {\beta})=\frac{\sum_{\beta} (v_{\alpha}(\beta)-\langle v_{\alpha} \rangle)(v_{\alpha}(\beta + \Delta \beta)-\langle v_{\alpha} \rangle )}{\sum_{\beta} (v_{\alpha}(\beta)-\langle v_{\alpha} \rangle)^2}.
\end{equation}
Where $\alpha$ defines the $x$, or $y$-velocity component. $\beta$ defines the $x$, or $y$-direction in which the correlation function is calculated and $\Delta {\beta}$ the spatial separation $ \Delta x$, or $ \Delta y$. The autocorrelation functions are a way to measure relative flow pattern constellation independent of the absolute spatial pattern position. Averaging these correlation functions in time and along the $x$-direction reveals stream patterns in the shear flow that occur at a high probability and are characteristic throughout the experiments. The 10000 frames at 2000 fps that are recorded allow to calculate 10000 space averaged autocorrelation functions at a time difference of 1/2000 seconds. To reduce the noise and strengthen the signal once more the autocorrelation functions are averaged over the total recorded time. The lifetime of a vortex must be longer than 1/2000 seconds to be detected which is the case for the discussed vortices. Figs. \ref{FVA1}, and \ref{FVA2} show this averaged autocorrelation functions of the $v_x$ and the $v_y$-component in set A. Below these plots Figs. \ref{FVA3}, and \ref{FVA4} show equivalent autocorrelation functions for the experimental set B. 
The plots \ref{FVA1}, and \ref{FVA3} show that the $v_x$-velocity component is maximally anticorrelated over a typical distance of $\Delta y\approx\,$4\,mm in $y$-direction. We notice that the anticorrelation of the $v_x$-velocity component is stronger in the middle of the film (see Fig. \ref{FVA1}) than close to the front film (see Fig. \ref{FVA3}). In contrast the $v_y$-velocity component is stronger anticorrelated close to the film (see Fig. \ref{FVA4}) than in the middle between the two sides of the film (see Fig. \ref{FVA2}). This is consistent with the flow snapshots of the $v_{y_c}$-velocity from experimental set A which previous displayed isotropic and random flow patterns in Figs. \ref{w2} and \ref{w3} in the middle of the cell. The corresponding autocorrelation functions in Fig. \ref{FVA2} also show almost no anticorrelation at $\Delta y\approx\,$4\,mm. This means that there are only very weakly correlated patterns of the $v_{y_c}$-velocity component for the time and space average in the middle of the cell. Instead a rather random pattern distributions must be the case in the turbulent situation.\\
Furthermore the maximal anticorrelation of the $v_y$-velocity component in Fig. \ref{FVA4} is located at the same spatial separation in $y$-direction as for the $v_x$-velocity component at $\Delta y\approx\,$4\,mm as shown in Figs. \ref{FVA1} and \ref{FVA3}. The maximal anticorrelation of the $v_y$-velocity component and the $v_x$-velocity component at the same spatial separation shows that neighboring layers preferably rotate counter-wise around the stream-wise direction. This type of movement resembles a roll with streamlines following a helix with an axis in $x$-direction. In the turbulent regime the autocorrelation functions show an increased probability that multiple helices coexist at the same time in the cell. The axis direction alternates in $x$-direction between neighboring helices while the chirality of these coexisting helices is the same. \\
The length of the rolls in horizontal stream-wise direction decreases at higher Reynold numbers. The autocorrelation of the $v_x$-velocity component as function of the horizontal stream-wise spatial separation $\Delta x$ is shown in Fig. \ref{FVA5a} for experiment set A and in Fig. \ref{FVA6a} for experiment set B. In both sets the horizontal $v_x$-velocity component is correlated over a shorter spatial separation with increasing Reynold numbers. \\ 
To understand the shape of these rolls velocity histogram are shown in Figs. \ref{FVA5} and \ref{FVA6} for experiment set A in the center of the sheer flow and in Figs. \ref{FVA8} and \ref{FVA9} for experiment set B close to the front film. Comparing the four plots shows that the $v_x$-component is distributed over a wider range compared to the $v_y$-component. This shows that the fluid flows faster horizontal along the positive or negative stream-wise $x$-direction than in span-wise $y$-direction. The main flow direction of a vortex is therefore the stream-wise $x$-direction in the $xy$-plane. \\
In the center of the film (Figs. \ref{FVA5}, and \ref{FVA6}) the $v_{x_c}$ and $v_{y_c}$-velocity component is distributed almost symmetric with respect to the $y$-axis. The same holds for the $v_{y_f}$-component close to the front film, which also follows an almost symmetric distribution with respect to the $y$-axis in Fig. \ref{FVA9}. Slight asymmetries in these distributions result from the fact that if particles inside the 1mm thick laser sheet are located towards the camera side they will be easier detected and recorded than particles on the other side of the laser sheet. Furthermore it is possible that the laser sheet position is marginally shifted from the middle of the cell. \\
Nevertheless compared to these rather symmetric distributions the $v_{x_f}$-component close to the front film loses this symmetric distribution and Fig. \ref{FVA8} shows that here the fluid is more likely to move along the direction the film is moving than against the film. In this case the film was moving in the negative $x$-direction where the $v_{x_f}$-ditribution also shows a higher probability compared to the movement in positive $x$-direction.\\ 
The previous results from the autocorrelation functions (see Figs. \ref{FVA1}, \ref{FVA2}, \ref{FVA3}, and \ref{FVA4}) show that turbulent rolls occupy the space between the two sides of the transparent band. The symmetric $v_{x_c}$-velocity distribution in the middle of the film means that roll shaped vortices move in the positive $x$-direction with the same probability as they move in the negative $x$-direction. Close to front film the situation however changes. Here the asymmetric distributions show that rolls moving against the direction of the front film have a lower $v_{x_f}$-velocity than rolls that moves in the same direction as the front film. From this we can conclude that the $v_x$-velocity of a turbulent roll is different on each side of the film and depends on the relative motion between the roll and the respective film side. The stream-wise $v_x$-velocity of a roll shaped vortex must therefore be a function of $z$ along the wall normal direction. \\ 
\begin{figure}
  \centerline{
  \subfigure[]{
    \epsfig{height=.4\textwidth,file=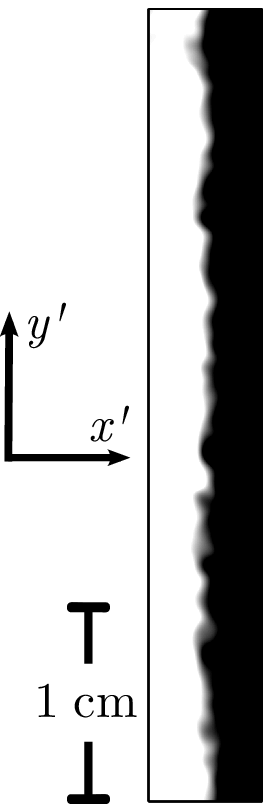}
    \label{w11}
  }
 \subfigure[]{
    \epsfig{height=.4\textwidth,file=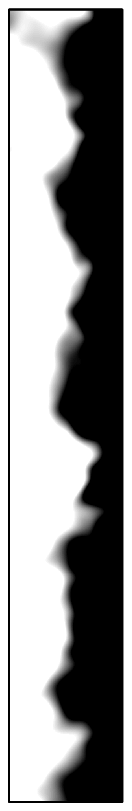}
    \label{w12}
  }
  \subfigure[]{
    \epsfig{height=.4\textwidth,file=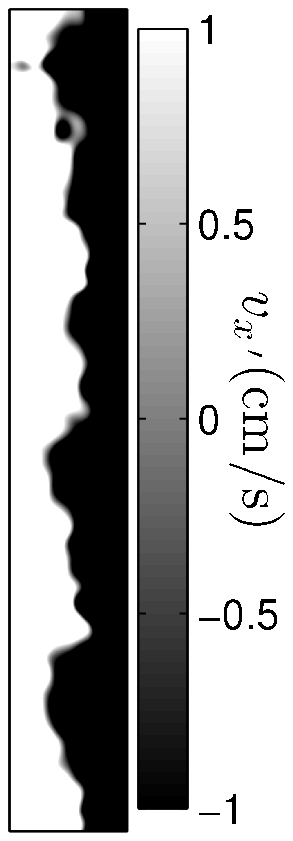}
    \label{w13}
   }}  
  \caption[Optional caption for list of figures]{Snapshots of the horizontal $v_{x^{\prime}}$-velocity component in the 52$^{\circ}$ view. The displayed area shows the whole cross section of the film and has a dimension of 0.55 cm by 4 cm. Reynolds numbers are Re=300 (laminar) in \subref{w11}, 400 in \subref{w12} and 1120 in \subref{w13}. A clear zero velocity line is visible at $x^{\prime}\approx 3$\,mm which is fluctuating for the two turbulent Reynolds numbers in plot \subref{w12} and \subref{w13}. To enhance the contrast and visualize the flow direction velocities are represented by a linear gray scale only from -1cm/s (black) up to 1cm/s (white) although the actual range of velocity values is much larger. Values below or above this linear gray scale are represented black or white respectively. For smother pictures the spatial resolution was increased by linear interpolation.}
  \label{wall1}
\end{figure}
\begin{figure}
  \centerline{
  \subfigure[]{
    \epsfig{height=.4\textwidth,file=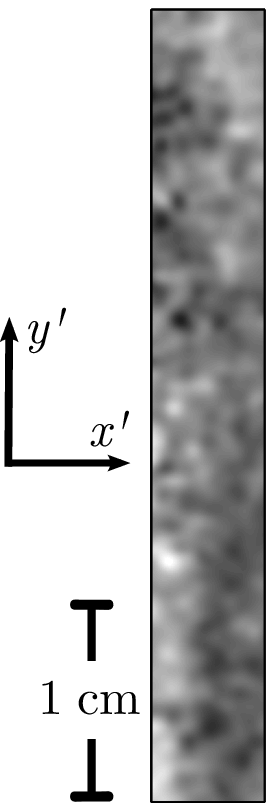}
    \label{w21}
  }
 \subfigure[]{
    \epsfig{height=.4\textwidth,file=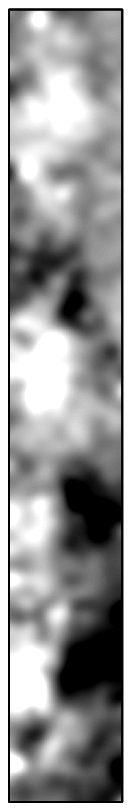}
    \label{w22}
  }
  \subfigure[]{
    \epsfig{height=.4\textwidth,file=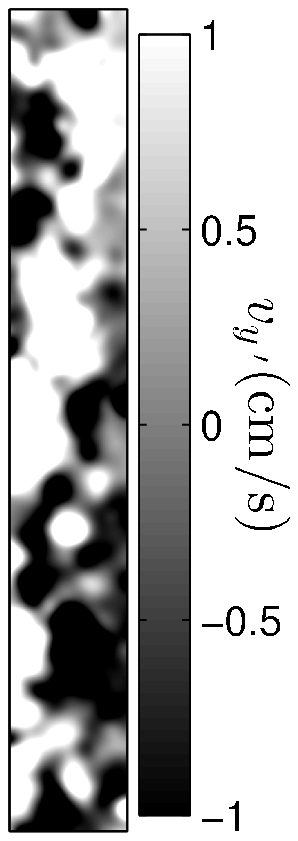}
    \label{w23}
   }}  
  \caption[Optional caption for list of figures]{Corresponding to Fig. \ref{wall1}. Snapshots of the horizontal $v_{y^{\prime}}$-velocity component in the 52$^{\circ}$ view. The displayed area shows the whole cross section of the film and has a dimension of 0.55\,cm x 4 cm. Reynolds numbers are Re=300 (laminar) in \subref{w21}, 400 in \subref{w22} and 1120 in \subref{w23}. To enhance the contrast and visualize the flow direction velocities are represented by a linear gray scale only from -1cm/s (black) up to 1cm/s (white) although the actual range of velocity values is much larger. Values below or above this linear scale are represented black or white respectively. For smother pictures the spatial resolution was increased by linear interpolation.}
  \label{wall2}
\end{figure}


\subsection{Inclined view at $\alpha=52^{\circ}$}
\begin{figure*}
  \centerline{
  \subfigure[]{
    \epsfig{height=.34\textwidth,file=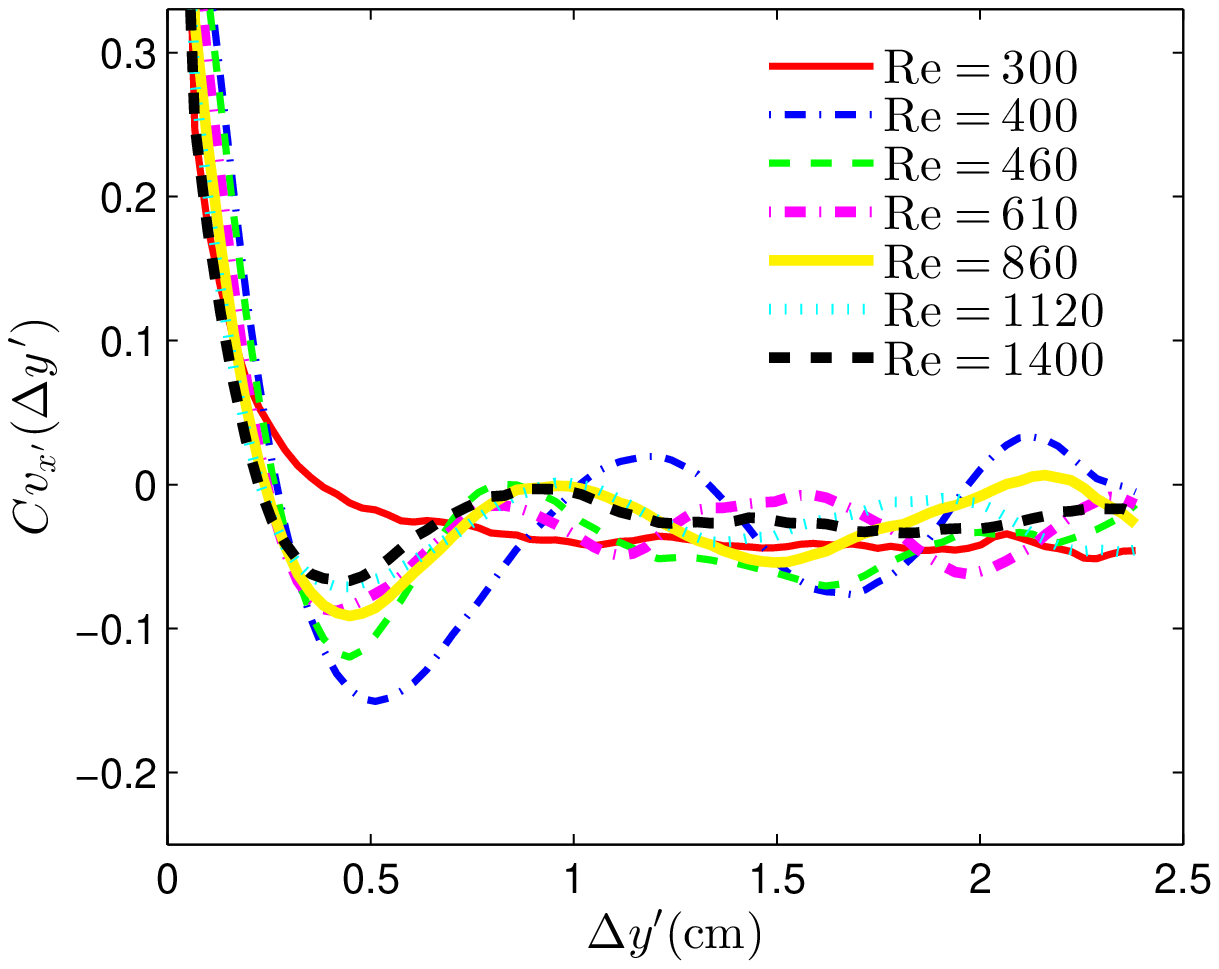}
    \label{FVA13}
  }
 \subfigure[]{
   \epsfig{height=.34\textwidth,file=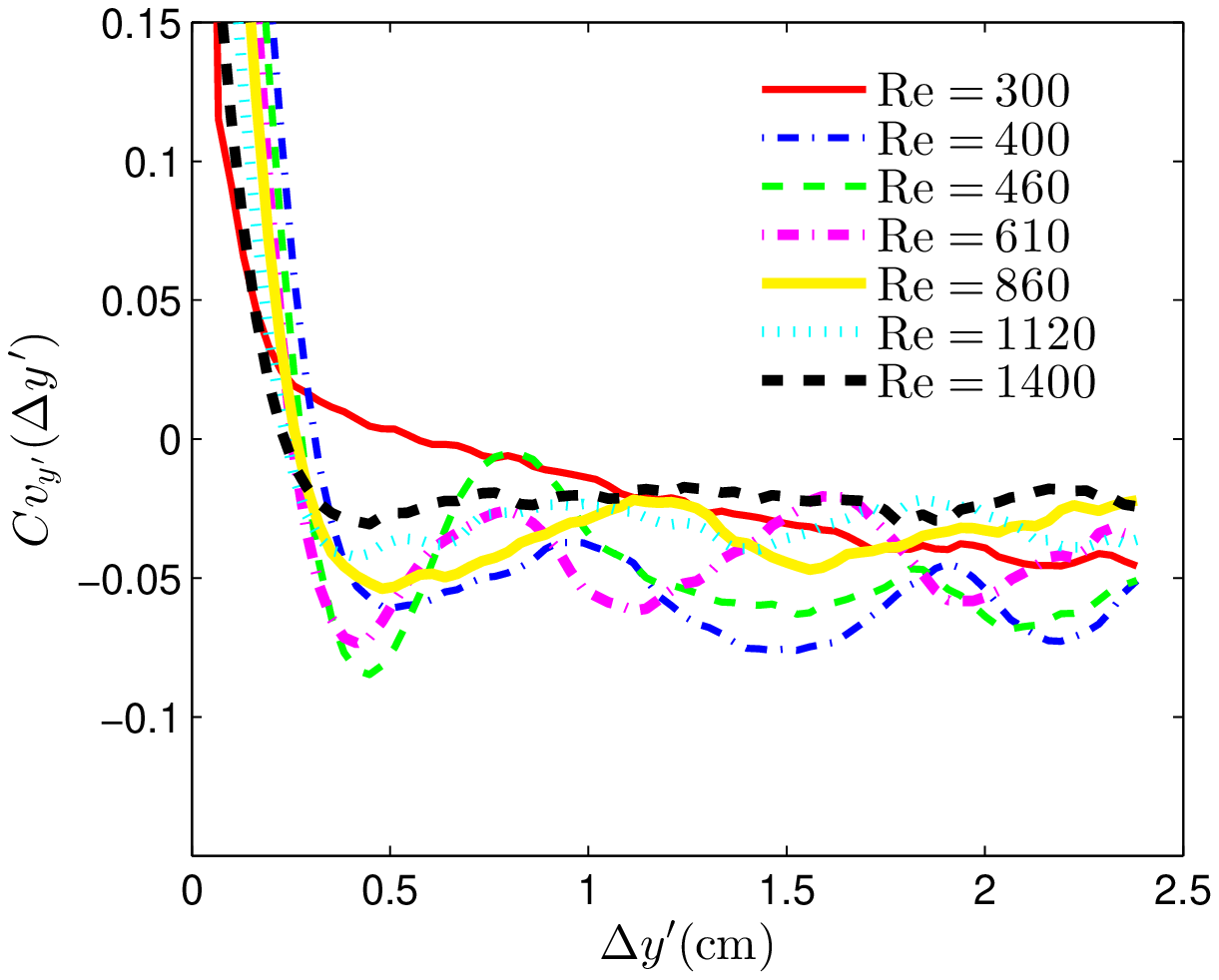}
    \label{FVA15}
  }}
\centerline{
 \subfigure[]{
   \epsfig{height=.34\textwidth,file=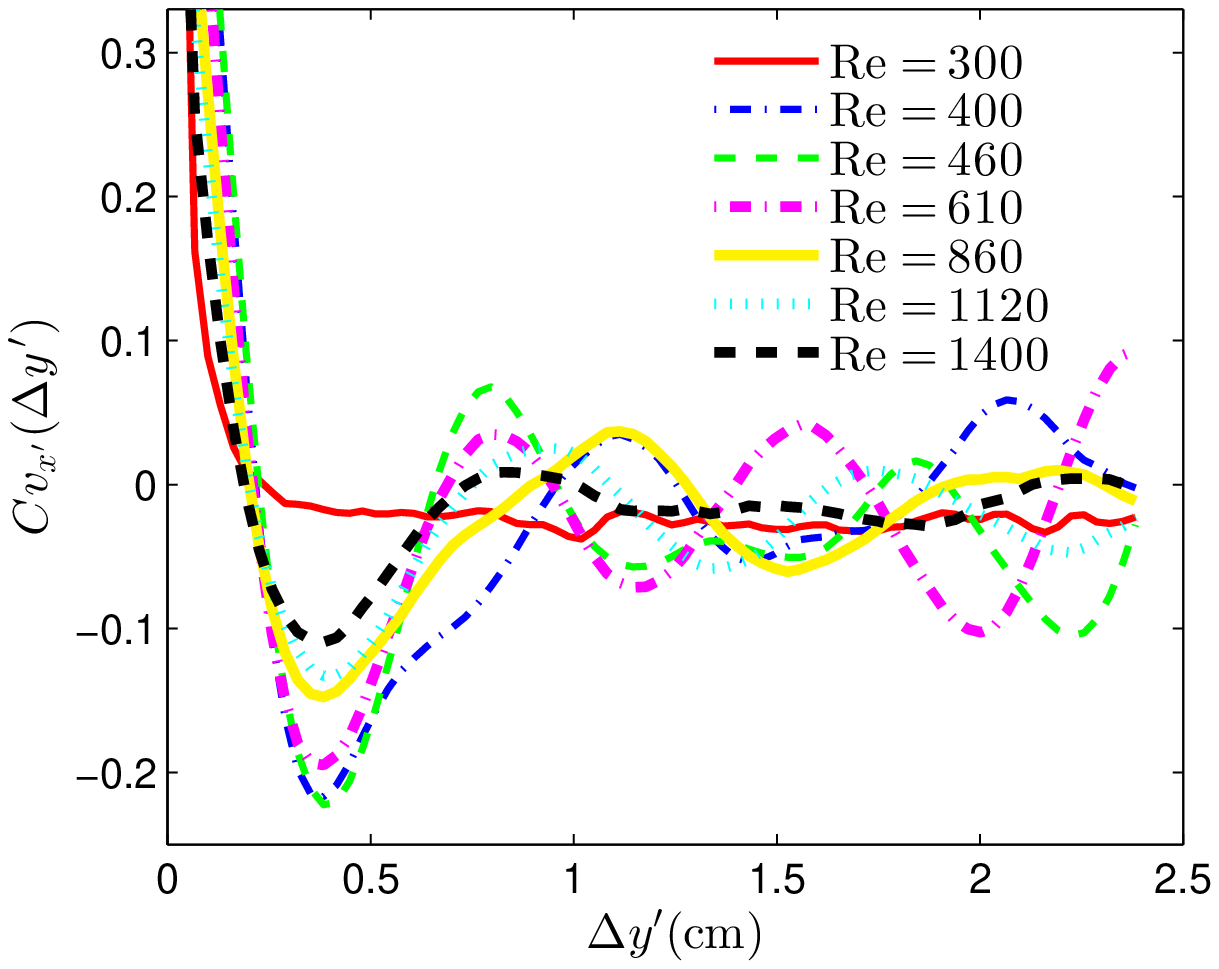}
    \label{FVA14}
  }
 \subfigure[]{
   \epsfig{height=.34\textwidth,file=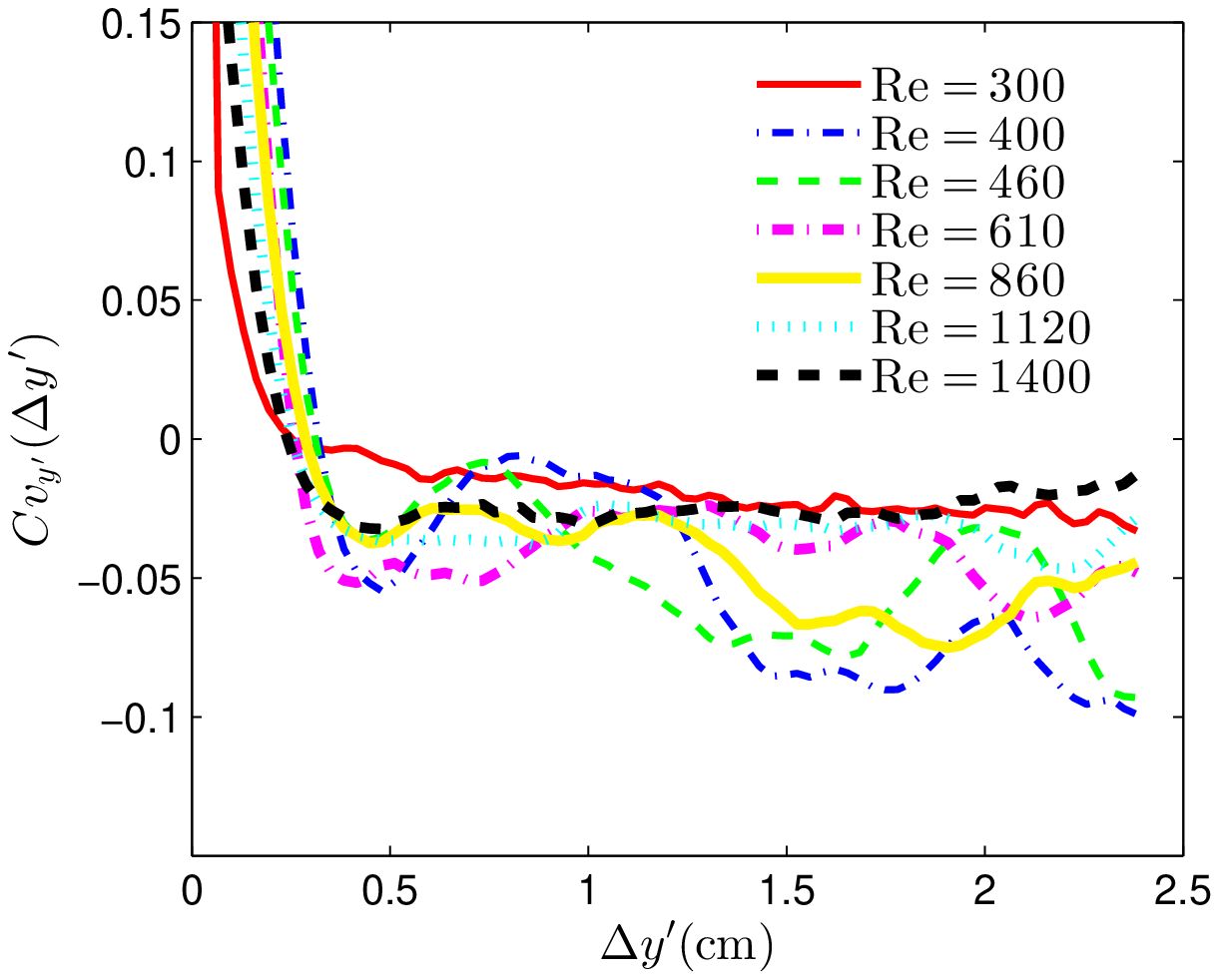}
    \label{FVA16}
  }}  
  \caption[Optional caption for list of figures]{Figs. \subref{FVA13} and \subref{FVA14} show autocorrelation functions for the $v_{x^{\prime}}$-velocity component and  Figs. \subref{FVA15} and  \subref{FVA16} show autocorrelation functions for the $v_{y^{\prime}}$-velocity component as a function of $\Delta y^{\prime}$. \subref{FVA13} and \subref{FVA15} are measured at $x^{\prime}= 1$\,mm away from the back film. \subref{FVA14} and \subref{FVA16} are measured at $x^{\prime}=3$\,mm in the middle between the two films.}
  \label{FVAall6}
\end{figure*}
\begin{figure*}
  \centerline{
  \subfigure[]{
    \epsfig{height=.34\textwidth,file=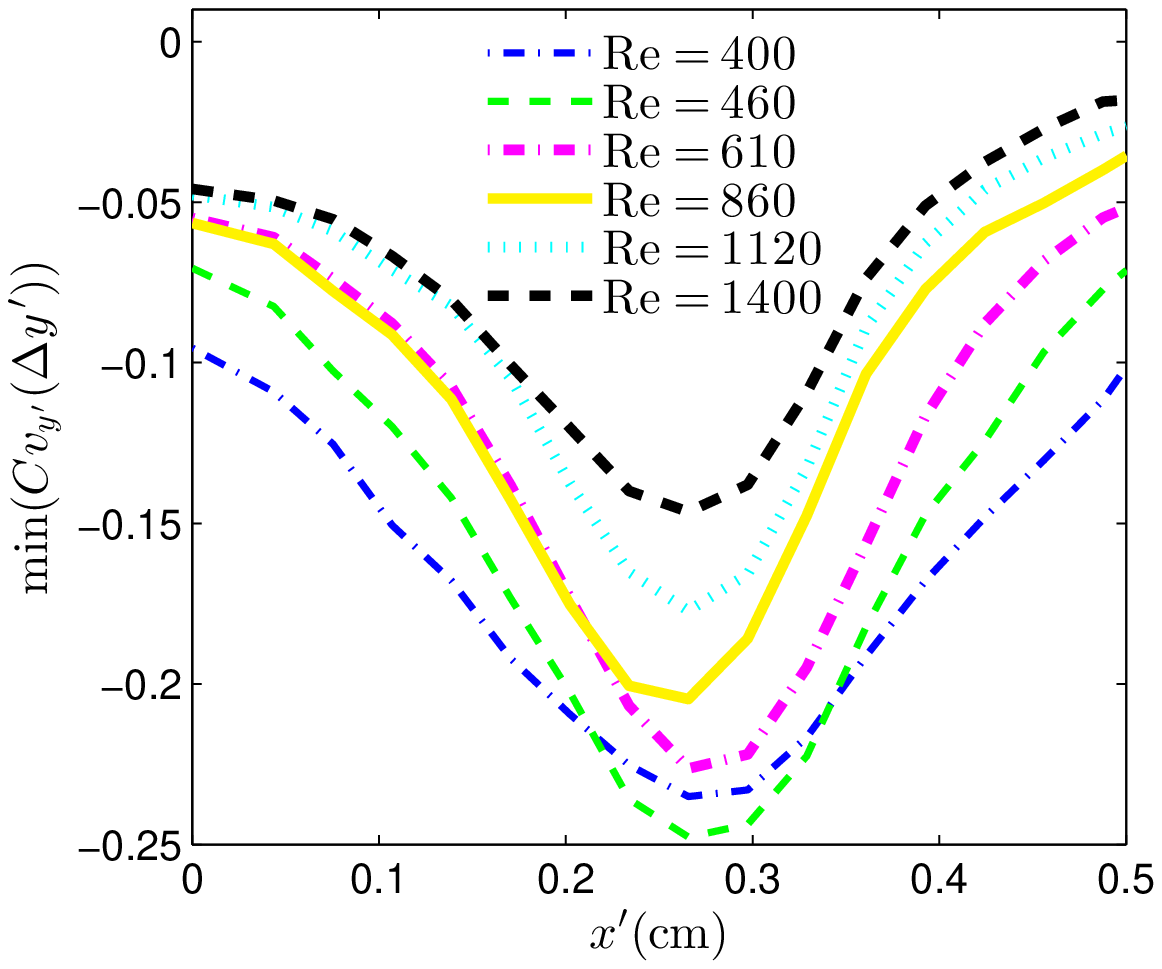}
    \label{FVA17}
  }
 \subfigure[]{
   \epsfig{height=.34\textwidth,file=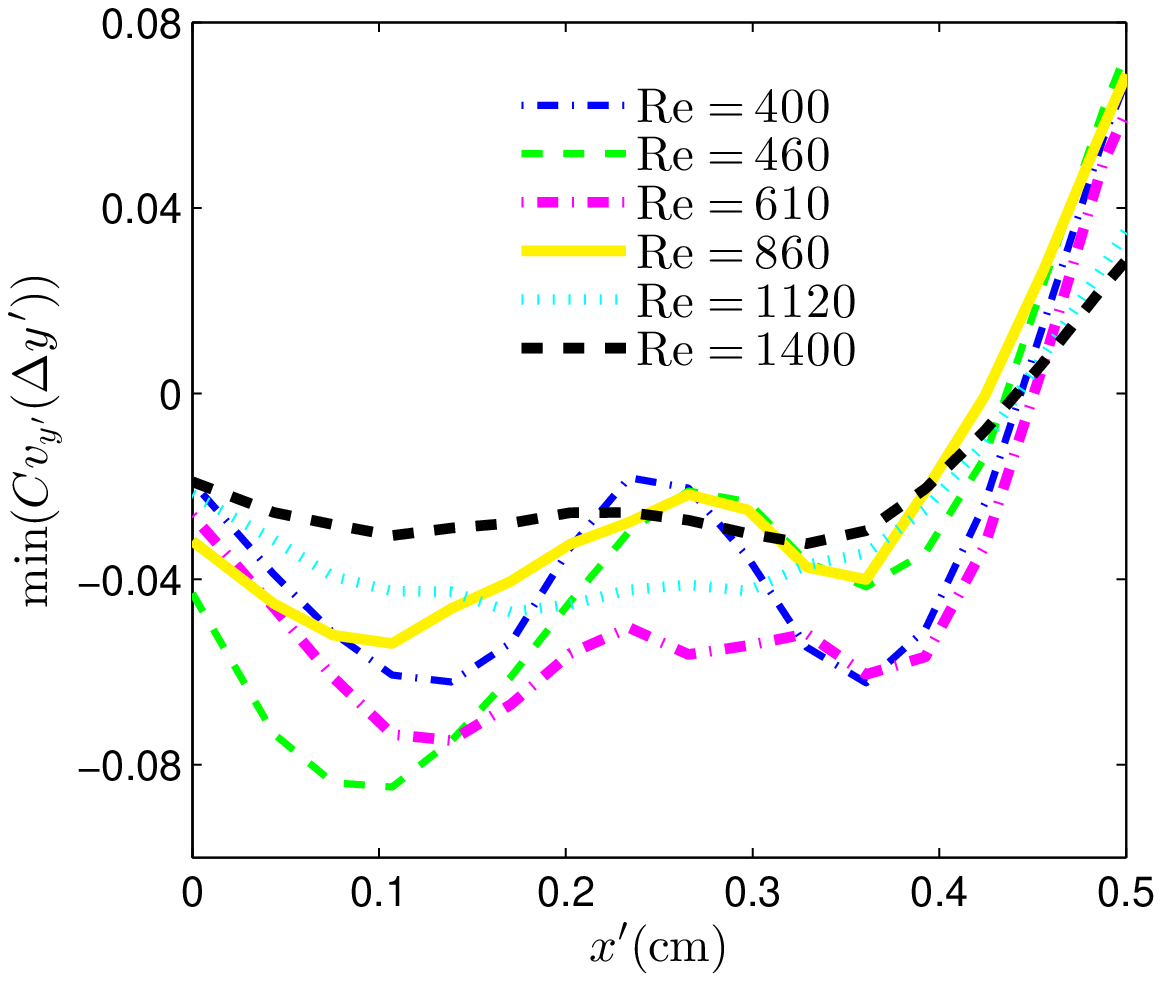}
    \label{FVA18}
  }}
  \caption[Optional caption for list of figures]{At several $x^{\prime}$-positions velocity autocorrelations as a function of $ \Delta y^{\prime}$ have been calculated. All of these functions have a first local minimum for increasing $\Delta y^{\prime}$. This value of the first maximum velocity anticorrelation is plotted for the tested $x^{\prime}$-positions. This is shown in plot \subref{FVA17} for the $v_{x^{\prime}}$-velocity component and in \subref{FVA18} for the $v_{y^{\prime}}$-velocity component. The laminar flow experiment at a Reynolds number of $\text{Re}=300$ does not show a consistent local minimum in the velocity autocorrelation functions and is therefore not part of the plots.}
  \label{FVAall7}
\end{figure*}
In the experimental setup previously referred to as set C the laser sheet enters the cell under an angle of 52$^\circ$. The angle has been calculated by Snell's law and refraction at the plexiglas wall and the refraction of the beam entering the the water has been taken into account. The setup allows to measure a projection of the velocity vector ${\bf v}$ on the $x^\prime y^\prime$-plane at an angle of $\alpha=$52$^\circ$ as shown in Fig. \ref{fig:expsetup} and stated in Eq. (\ref{eq:coordtrans}). In this way the projection of the wall normal flow direction can be studied although it is important to state that the recorded $v_{x^\prime}$-component depends on both the old $v_x$ and $v_z$-velocity components. The particles tracked at each position by the CIV method are randomly located within the finite thickness of the laser sheet. Between the two extreme positions on both sides of the laser sheet the maximal possible velocity variation are $\Delta v_{x^{\prime}}=\sin(52^{\circ})\cos(52^{\circ}) L_d U/d=m=\pm 0.1 U $ but this will average to zero in time. From the recorded pictures the flow field was again calculated by a correlation image velocimetry CIV method. The resulting flow profile is shown in Fig. \ref{wall1} for the $v_{x^{\prime}}$-component, in Fig. \ref{wall2} for the $v_{y^{\prime}}$-component. The shown velocity field in these pictures is linearely interpolated between CIV measurement points. The three snapshots correspond to the Reynolds numbers Re=300 in Figs. \ref{w11} and \ref{w21} (laminar), Re=400 in Figs. \ref{w12} and \ref{w22} and Re=1120 in Figs. \ref{w13} and \ref{w23}. The gray scale intends to highlight the direction of the flow rather than to quantify the velocity. Looking at Figs. \ref{wall1} 
this gray scale makes it easier to illustrate that a clear zero velocity line in $y^{\prime}$-direction appears between the two flow directions in the middle between the two sides of the film at $x^{\prime}\approx 3$\,mm. For the laminar situation this interface of zero velocity is a almost straight line in Fig. \ref{w11} (laminar). When the band velocity is increased and the flow becomes turbulent, the zero velocity line starts to curve and fluctuate along the $y^{\prime}$-direction at $x^{\prime}\approx 3$\,mm as shown in Figs. \ref{w12} and \ref{w13} 
These result correspond well to the analytical work of \cite{clever}. Concerning the $v_{y^{\prime}}$-velocity component in Fig. \ref{wall2} it is not possible to draw systematic conclusions from the snapshots apart from the impression that the turbulent flow in Figs. \ref{w22} and \ref{w23} displays random patterns that appear larger than in the laminar situation in Fig. \ref{w21}. For a better insight once more velocity autocorrelation functions are used as defined in Eq. (\ref{autoc}). The autocorrelation function in the $y^{\prime}$-direction in this setup must not be averaged along the $x^{\prime}$-direction because of the heterogeneity of the flow in this direction. For this reason the autocorrelation function in the $y^{\prime}$-direction is calculated at each $x^{\prime}$ value separately and only averaged along $y^{\prime}$ and in time. These results are presented in Fig. \ref{FVAall6}. Figs. \ref{FVA13} and \ref{FVA14} show autocorrelation functions for the $v_{x^{\prime}}$-velocity component and Figs. \ref{FVA15} and \ref{FVA16} show autocorrelation functions for the $v_{y^{\prime}}$-velocity component.\\
 The autocorrelation functions in Figs. \ref{FVA13} and \ref{FVA15} are measured at $x^{\prime}= 1$\,mm. This is close to the film side that lies furthest away from the the camera. The plots in Figs. \ref{FVA14} and \ref{FVA16} are measured at $x^{\prime}= 3$\,mm in the middle between the two films. We notice that the $v_{x^{\prime}}$-velocity component is significantly stronger anticorrelated in the middle (see Fig. \ref{FVA14}) than at the proximity of the film in Fig. \ref{FVA13}. Interestingly the opposite observation can be made regarding the autocorrelation functions of the $v_{y^{\prime}}$-velocity component which is stronger anticorrelated at the proximity of the film Fig. \ref{FVA15} than in the middle Fig. \ref{FVA16}. This again is consistent with the picture of counter-wise rotating rolls which are moving in opposite horizontal directions. \\
Apart from the two displayed velocity autocorrelation functions additional velocity autocorrelations as a function of $\Delta y^{\prime}$ have been calculated at several $x^{\prime}$-positions. In total such autocorrelation functions were calculated at 19 $x^{\prime}$-positions. The main difference between these functions is the value of the first local minimum where the anticorrelation becomes maximal for the first time while $\Delta y^{\prime}$ is increased. In Fig. \ref{FVA17} we show how the value of this local minimum evolves for the tested $x^{\prime}$-positions for the $v_{x^{\prime}}$-velocity component and in Fig. \ref{FVA18} for the $v_{y^{\prime}}$-velocity component. The laminar experiment is not shown in these plots due to the lack of a local minimum in the autcorrelation functions.\\
As expected the $v_{x^{\prime}}$-velocity component has the strongest anticorrelation in the middle between the film at $x^{\prime}\approx 3$\,mm. Unlike the plot for $v_{y^{\prime}}$-velocity component which clearly displays two minima in Fig. \ref{FVA18} both located at the sides separated by a local maximum in the middle of the film. The two $x^{\prime}$-positions where the maximal anticorrelation is found are $x^{\prime}\approx 1$\,mm and $x^{\prime}\approx 3.9$\,mm. For increasing $x^{\prime}$ values the measurement of the velocity was interfered by reflections from the front film and the front plates of the cell where the laser beam first hit the cell. For this reason velocity values are not symmetric around $x^{\prime}\approx 3$\,mm. \\
In all plots it could also been observed that in the turbulent regime at higher Reynolds numbers the anticorrelation of the flow decreased. This showed that at higher fluid velocity the flow patterns become more random and the probability of ordered patterns gradually decreases.
\section{Conclusion:}
\label{Conclusions}
\begin{figure}[ht!]
  \includegraphics[width=1.0\columnwidth]{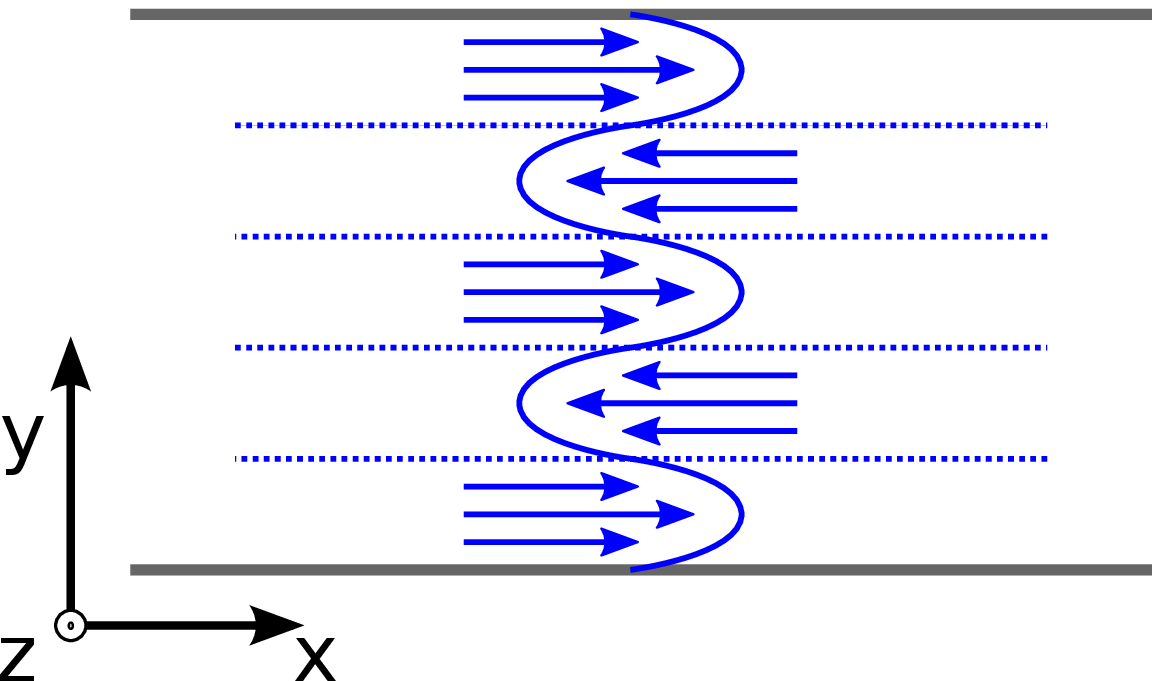} 
  \caption{\label{scmF}(Color online) A front view of the cell taken at the center of the model corresponding to experimental set A. The flow is shown in the preferred however idealized configuration.}
\end{figure}
\begin{figure}
  \centerline{
  \subfigure[]{
    \epsfig{height=.20\textwidth,file=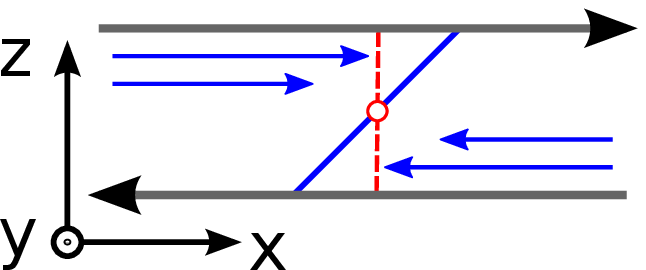}
    \label{scm1}
  }}
\centerline{
 \subfigure[]{
   \epsfig{height=.16\textwidth,file=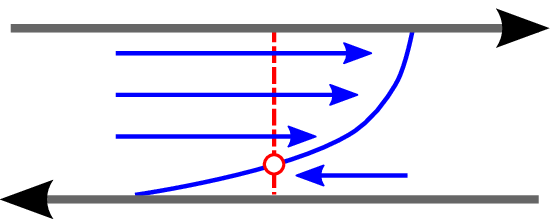}
    \label{scm2}
  }}
\centerline{
 \subfigure[]{
   \epsfig{height=.16\textwidth,file=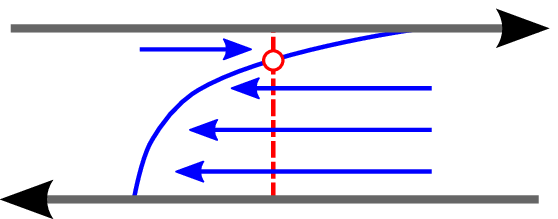}
    \label{scm3}
  }}
  \caption[Optional caption for list of figures]{(Color online) Idealized preferred flow configurations are shown from the top for a cross section in the horizontal $x$,$z$-plane. In \subref{scm1} the typical laminar flow is shown and marked with the (red)-circle where the fluid has zero $x$-velocity in the middle between the film sides. In \subref{scm2} and in \subref{scm3} possible interpretations of horizontal roll cross sections are shown corresponding to a turbulent flow. The point in the flow of zero $x$-velocity is marked with a (red) circle. This point moves along the (red) dashed line when the $x$-flow direction of the roll is reversed.}
  \label{scmA}
\end{figure}
\begin{figure}[t!]
  \includegraphics[width=0.4\columnwidth]{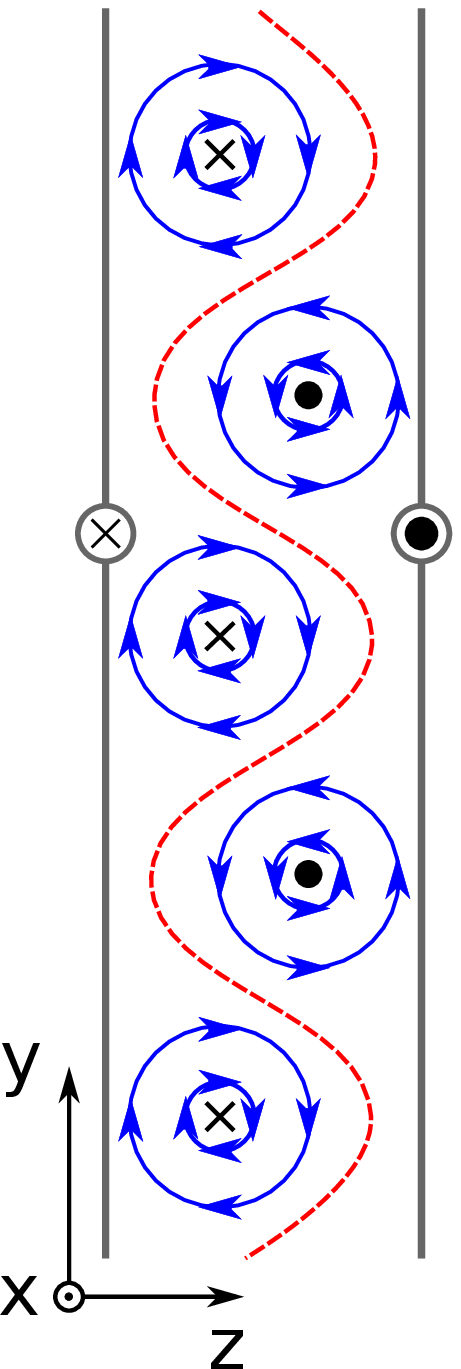} 
  \caption{\label{scmB} Several rolls are shown in an cross section along the $yz$-plane. The (red) dashed line between the roll vortices marks zero $x$-velocity. In this proposed idealized demonstration the rolls are assumed circular in the cross section along the $yz$-plane while in reality more complex shapes are likely.}
\end{figure}
The analysis of the velocity measurements showed a clear distinction between the laminar and turbulent flow at different Reynolds numbers. Further more the measurement allowed to quantify and qualify characteristic vortices that reappear at different times and positions during the experiments with an increased probability.
In the stream-wise $x$-direction such vortices are elongated and reach lengths of more than the observation area of 4\,cm. When the Reynold number is increased the stream-wise length decreases. In the $y$ and $z$-direction the size of the vortices is sharply confined to 4\,mm which corresponds to the film separation as shown in the idealized Fig. \ref{scmF}. Typically several such vortices are found in the cell at the same time. The vortices are correlated in their relative position and velocity to each other. In $y$-direction these vortices are separated by 4\,mm. Many such vortices can be stacked on top of each other in the $xy$-plane. The main direction of fluid movement in each vortex is the $x$-direction. Neighboring vortices move in the opposite $x$-direction. \\
In Fig. \ref{scmA} cross sections in the $xz$-plane of idealized flow patterns are shown in top view. In Fig. \ref{scm1} a typical laminar shear flow is illustrated. The fluid has a linear velocity profile in $z$-direction with zero overall velocity in the middle marked with a (red) circle.\\
At higher Reynold numbers stream-wise elongated vortices have developed. From our previous analysis we can outline a number of flow field features of these vortices or turbulent rolls in more details. In a first step we could initially show that at the middle between the two film sides vortices have an either positive or negative main $x$-flow direction (see Figs. \ref{rall}, \ref{wall} and \ref{FVA5}, \ref{FVA6}). Here the main flow velocity of all vortices is equally distributed in positive and negative $x$-direction in a time and space average this is also shown in simulations \cite{cvitanovic, cvitanovic2}.\\ 
In a second step we studied how the main $x$-flow velocity inside a vortex is influenced by the film boundaries in $z$-direction. In $z$-direction each vortex is confined by the two sides of the film. There is always one side that moves along with the main $x$-flow direction of the vortex and the other film side that moves against the main $x$-flow direction of the vortex. When moving in $z$-direction from the middle between the film towards the film side that moves along with the main $x$-flow direction the $v_x$-velocity inside a vortex increases. When moving to the other side from the middle between the film towards the film side that moves against with the main $x$-flow direction the $v_x$-velocity inside a vortex decreases. This was shown in Fig. \ref{FVA8}. \\
At last we know that at the film boundary the fluid velocity must be the same as the film velocity. This condition together with the measurements taken can be combined to outline the flow profile of vortices in the $xz$-plane. A possible interpretation for the $v_x$ flow velocity inside turbulent vortices is given in Figs. \ref{scm2} and \ref{scm3}.\\
On one side of these vortices the film moves in the same stream-wise direction as the vortex while on the other side the film moves against the stream-wise flow of the vortex. As shown before when analyzing the velocity distributions (see Fig. \ref{FVA8}) the vortices display an asymmetry towards the two sides of the film and an overall shear of the vortices occurs. How this shear profile looks in detail when crossing the vortex in $z$-direction can not be stated exactly from the measurements discussed in this article. The precise profile shown in the figures is therefore a guess that fulfills the above discussed conditions. However the point of zero $x$-velocity is always located closer to the side of the film moving opposite to the vortex' main stream-wise flow direction as measured in Fig. \ref{FVA8} and shown in Figs. \ref{w12} and \ref{w13}. In Figs. \ref{scmA} this point is always marked with a (red) circle. \\
Moving along the $y$-direction the position of zero stream-wise flow velocity travels from one side of the film to the other between two neighboring vortices. In a cross section along the $yz$-plane zero stream-wise flow velocity is indicated in Fig. \ref{scmB} by a dashed (red) line. The snapshots of experimental set C show similar curved lines of zero $x'$-velocity in Figs. \ref{w12} and \ref{w13} which primarily show the $x$-velocity. \\
In addition to the horizontal flow in $x$-direction vortices also rotate around the $x$-axis which is consitence with among others \cite{Krug,Bottin} (experimental) and \cite{clever,Holstad, cvitanovic} (simulation). Neighboring vortices rotate counter-wise. As shown in the idealized Fig. \ref{scmB} by assuming circular shaped rolls in the $yz$-plane cross section. Combining the measured flow properties leads to a helix type of fluid motion around the $x$-axis in each of those vortices or rolls. \\
In summary a number of distinct features of typical vortices in a plane shear flow at high Reynolds number were revealed and discussed. An exact measurement of the shear and rotation of a typical vortex is considered to be a possible future step.\\
{\bf Acknowledgements:} We acknowledge the support of the Norwegian research council through the Frinat grant 205486 of NEEDS-MIPOR, the regional network REALISE, the ANR LANDQUAKE and the ITN FLOWTRANS. We thank Walter Goldburg, Eirik G. Flekk{\o}y, Luiza Angelutha, Joachim Bergli, Joachim Mathiesen, Mogens Jensen, Mickael Bourgoin and Romain Volk for fruitful discussions.

\end{document}